\documentclass[%
reprint,
nofootinbib,
 amsmath,amssymb,
 aps,
floatfix,
]{revtex4-2}

\usepackage{graphicx}
\usepackage{dcolumn}
\usepackage{bm}
\usepackage{mathrsfs}
\usepackage{subcaption}

\usepackage{multirow}
\usepackage{array}
\usepackage{hyperref}

\begin{document}

\title{
Ghost projection. II. Beam shaping using realistic spatially-random masks
}

\author{David Ceddia}
\affiliation{School of Physics and Astronomy, Monash University, Victoria 3800, Australia}
\author{Andrew M.~Kingston}
\affiliation{Department of Applied Mathematics, Research School of Physics and Engineering, The Australian National University, Canberra, Acton 2601, Australia}
\affiliation{CTLab: National Laboratory for Micro Computed-Tomography, Advanced Imaging Precinct,
The Australian National University, Canberra, Acton 2601, Australia}
\author{Daniele Pelliccia}
\affiliation{Instruments \& Data Tools Pty.~Ltd., Victoria 3178, Australia}
\author{Alexander Rack}
\affiliation{ESRF -- The European Synchrotron, CS40220, 38043, Grenoble, France}
\author{David M.~Paganin}
\affiliation{School of Physics and Astronomy, Monash University, Victoria 3800, Australia~}

\date{\today}

\begin{abstract}
Spatial light modulation is important for many scientific and industrial applications. The spatial light modulator and optical data projector both rely on precisely configurable optical elements to shape a light beam.  Here we explore an image-projection approach which does not require a configurable beam-shaping element. We term this approach {\em ghost projection} on account of its conceptual relation to computational ghost imaging. Instead of a configurable beam shaping element, the method transversely displaces a single illuminated mask, such as a spatially-random screen, to create specified distributions of radiant exposure.  The method has potential applicability to image projection employing a variety of radiation and matter wave fields, such as hard x rays, neutrons, muons, atomic beams and molecular beams. Building on our previous theoretical and computational studies, we here seek to understand the effects, sensitivity, and tolerance of some key experimental limitations of the method.  Focusing on the case of hard x rays, we employ experimentally acquired masks to numerically study the deleterious effects of photon shot noise, inaccuracies in random-mask exposure time, and inaccuracies in mask positioning, as well as adapting to spatially non-uniform illumination.  Understanding the influence of these factors will assist in optimizing experimental design and work towards achieving ghost projection in practice.  
\end{abstract}

\maketitle


\section{Introduction}

Speaking of ``building signals out of noise'' is not necessarily a contradiction in terms. This notion underpins both classical ghost imaging \cite{shapiro2008computational,Bromberg2009,Katz2009} and random-basis decomposition \cite{Gorban2016}, which leads to the related idea of {\em ghost projection} \cite{ceddia2022ghost}. The key concept is that an ensemble of spatially-random masks may be illuminated, with  uniform or non-uniform exposure times, to generate an arbitrary desired spatial distribution of radiant exposure \cite{paganin2019writing,ceddia2022ghost}.  The arbitrariness of the resulting spatial distribution is limited by the highest spatial frequencies in the illuminated masks, and a constant additive offset or pedestal.  

The ensemble of illuminated random masks may be generated by transversely scanning a single random mask \cite{paganin2019writing}. A single non-random mask may also be employed, e.g.~using a mask fabricated to be orthogonal under transverse translation \cite{Svalbe2020}. Interestingly, as we shall argue, there are circumstances in which random masks may be more efficient than orthogonal masks, for ghost projection.

Ghost projection may be viewed as a reversed form of classical computational ghost imaging (CCGI). In particular, CCGI \cite{shapiro2008computational} reconstructs an unknown sample transmission function by sequentially interrogating that sample using a set of illuminated masks, and then collecting the total transmission (bucket signal) for each mask. A reflection geometry is also possible, but for clarity we henceforth only refer to the transmission case. The bucket signals can be correlated with the known masks, to reconstruct the transmission function of the sample. Conversely, ghost projection creates a desired time-integrated transmission pattern by sequentially illuminating a set of suitable masks \cite{ceddia2022ghost,paganin2019writing}. Here, the mask exposure times are analogous to bucket signals. Thus, rather than bucket signals being measured to determine an unknown transmission function in CCGI, a desired known time-integrated transmission function is shaped by specifying mask exposure times in ghost projection.

The ghost-projection concept sidesteps the need to have a precisely configurable dynamic beam-shaping element. Configurable elements are the key concept underlying spatial light modulators (SLMs) and optical data projectors in the visible-light regime, but also the key limitation to extending SLMs and data projectors to other regimes \cite{Shroff2001,Chkhalo2017,Olofsson2021}. Along similar lines, ghost projection has no need for a precisely configured static beam-shaping element (or elements), which is the main idea behind mask-based projection lithography. The above examples suggest that considerable flexibility might arise, from not needing a precisely configurable---or precisely configured---beam-shaping element. In particular, ghost projection opens the possibility of data projectors using matter and radiation wave fields for which such projectors do not exist, or pushing mask-based lithography into such regimes.  Fields, for which ghost projection might in future be employed, include hard x rays \cite{Paganin2006}, gamma rays, muon beams \cite{MuonRadiography, Yamamoto2020}, molecular beams \cite{MolcularBeamsBook}, atom beams \cite{AtomBeamBook}, atomic lasers \cite{AtomLasers2013}, and neutron beams \cite{NeutronOpticsHandbook}.   

It is worth briefly comparing some of the ideas in the preceding paragraph, with what has already been achieved in pushing lithography and configurable spatial modulators beyond the visible-light and extreme-ultraviolet realms.  X-ray lithography requires one to account for the proximity effect \cite{Bourdillon2000,Bourdillon2001}, whereby the effects of coherent free-space diffraction---between the exit surface of the mask and the entrance surface of the projection substrate---cannot be neglected in mask design. The problem becomes one of inverse inline holography \cite{Gabor1948}, with the lithographic mask needing to be the inline hologram that leads to a specified distribution of radiant exposure over the substrate. A number of schemes, in which masks are designed to contain inbuilt compensation for the proximity effect, have shown significant promise in short-wavelength photon lithography \cite{Bourdillon2000,Bourdillon2001}.  Interference lithography has also been pursued with some success in short-wavelength regimes \cite{Zhao2020}.  High-resolution mask-based lithography and configurable modulators in the neutron, ion, molecular-beam and atomic-beam regimes, by way of comparison, appear to be less well developed. Similarly, while impressive advances continue to be made in configurable-spatial-modulator technology \cite{Chkhalo2017,Olofsson2021}, it remains challenging to adapt the existing technology beyond the visible-light-regime, for example to the extreme ultraviolet and shorter-wavelength domains.  Atom-beam lithography has been pursued using optical and magnetic beam-shaping \cite{AtomLitho1,AtomLitho2}. In the context of the single-configured-mask lithographic or configurable-mask spatial-modulator strategies indicated here, ghost projection may be viewed as an alternative whose potential utility is worth exploring, in our view, in the regimes listed at the end of the previous paragraph.

The longer-term objective of this work is to practically demonstrate ghost projection with radiation for which no current straightforward projection mechanism exists. In initial steps towards this goal, we consider hard x rays as an indicative example.  In recent papers we explored a range of techniques (combinations of randomly-patterned illuminations and exposure times) to achieve ghost projection, and determined the most effective to be via numerical optimization \cite{paganin2019writing,ceddia2022ghost}. The present paper represents a next step towards the longer-term goal.  Having determined the method to combine patterned illuminations to achieve ghost projection in theory, here we explore some of the main experimental considerations that will affect performance. Understanding the effects, sensitivity, and tolerance of each experimental limitation will enable the optimization of experimental design in the future.

We close this introduction with a brief overview of the remainder of the paper.  Section II outlines the main concepts of ghost projection, by summarizing some key findings from our two preceding papers on this topic.  Section III considers two types of realistic spatially-random mask for hard x rays, namely metallic foam and sandpaper.  These masks are considered to be realistic because they are experimentally acquired, rather than being computationally modeled (as was the case in our previous papers). The former mask (metallic foam) has the contrast of its spatially-random bubble network enhanced by propagation-based phase contrast, i.e.~the sharpening effects of Fresnel diffraction in the slab of free space between the mask and the exposure plane \cite{Snigirev1995,Cloetens1996,Wilkins1996}.  The latter mask, namely the sandpaper, primarily employs attenuation contrast to generate its spatially-random pattern.  Having characterized the relevant statistical properties of the experimentally-obtained images of our two masks in Sec.~III, Sec.~IV studies from a computational perspective how these masks may be employed for x-ray ghost projection.  We focus, here, on several practical issues.  These issues include the relative efficacy of different schemes for ghost projection, the deleterious effects of shot-noise statistics, errors in the mask exposure time, and the deleterious effects of errors in transverse positioning. We also explore the ability of ghost projection to adapt to spatially non-uniform illumination, e.g.~a Gaussian source intensity profile. In all of these studies, both the ghost projection pedestal and signal-to-noise ratio are seen to give useful metrics for the performance of the ghost projections. The key outcome of this section is the development of several rules of thumb that we believe to be important for future experimental realizations of ghost projection.  We discuss some of the broader implications of our work, together with possible avenues for future research, in Sec.~V. Concluding remarks are made in Sec.~VI.     

\section{Ghost Projection}

\begin{figure}[ht!]
\centering
\includegraphics[width=0.85\columnwidth]{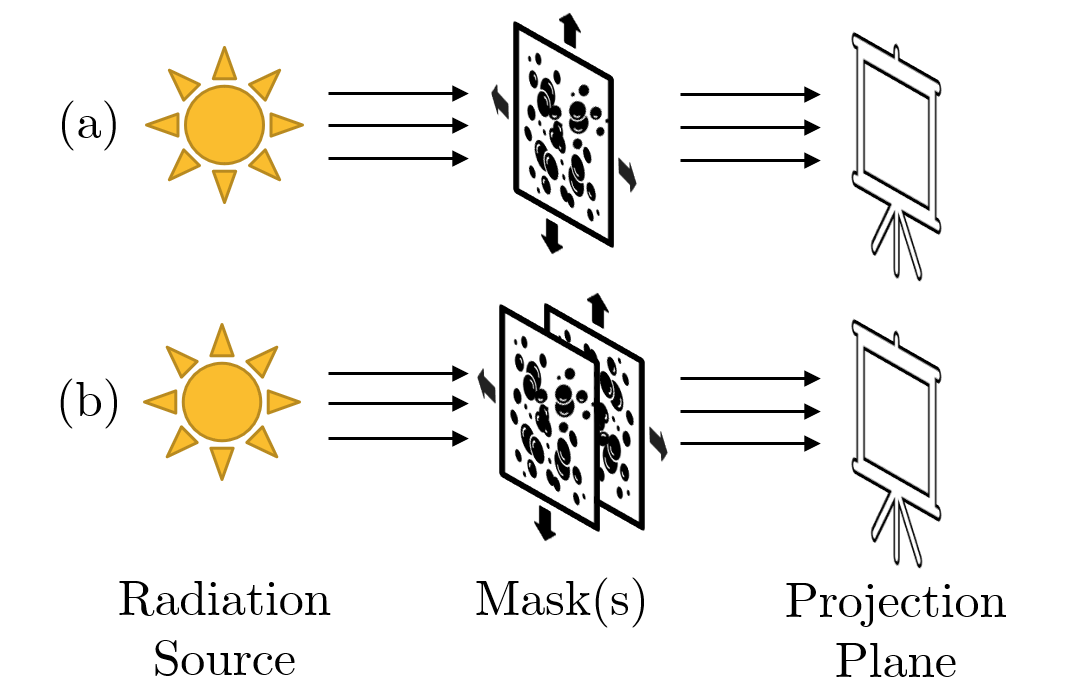}
\caption{Generic setup for ghost projection. (a) A radiation source illuminates a single mask, which may be transversely displaced to a number of specified positions in order to create a desired distribution of radiant exposure over the projection plane.  The mask may be spatially random, or manufactured to e.g.~yield patterns that are orthogonal under mask translation.  (b) Illustration of two or more independently translatable masks which may also be employed for ghost projection.  This exponentially increases the number of available configurations, compared to the case of a single mask.}
\label{Fig:GenericGhostProjectionSetup}
\end{figure}

A generic scheme for ghost projection is sketched in Fig.~\ref{Fig:GenericGhostProjectionSetup}a.  Here a radiation source illuminates a mask, which may be transversely displaced to a number of specified positions, in order to create a desired distribution of radiant exposure over the projection plane \cite{paganin2019writing,ceddia2022ghost}.  The mask may be spatially random, or manufactured to yield patterns that are orthogonal under mask translation.  The illumination time for each mask can be taken to be fixed, although a more general case is to allow the exposure times to be chosen differently for each transverse position of the ghost-projection mask. The key problem of ghost projection, for the single-mask scheme in Fig.~\ref{Fig:GenericGhostProjectionSetup}a, is to select the transverse mask locations (and, where applicable, the mask exposure times) to be such that a desired distribution $I$ of radiant exposure is created over the projection plane. The spatial distribution of radiant exposure is arbitrary, up to (i) a limiting spatial resolution that is dictated by the highest spatial frequencies present in the intensity distribution created by each mask over the projection plane, and (ii) an additive constant. In Fig.~\ref{Fig:GenericGhostProjectionSetup}b, we illustrate the possibility of having two or more independently translatable spatially-random masks.\footnote{As pointed out in footnote 2 of Ref.~\cite{ceddia2022ghost}, the idea of stacking multiple masks in the context of ghost projection is due to Kaye~S.~Morgan (Monash University, Australia).} This has the advantage of being linear in experimental requirements to image each master mask, but exponential in the number of configurations they can create.  Regarding this last-mentioned point, it is a general feature of ghost projection that, the larger the ensemble of linearly-independent masks from which a suitable subset can be chosen for illumination, the more efficient the ghost-projection can be.    

We now give a brief overview of the relevant formalism for ghost projection, as developed in Ref.~\cite{ceddia2022ghost}. Consider a discrete representation of the desired image that we wish to project.  We denote this discrete representation by $I_{ij}$, where $i \in [1,m]$ and $j \in [1,n]$ are integers that index the spatial extent of the image. For this investigation, we enforce the image to be zero-averaged (i.e.~E$[I]=0$, where E denotes expectation value) and enforce a contrast of unity. We define the ghost projection of this image as the following linear combination of random masks:
\begin{align} 
P_{ij} \equiv  R_{ij}^{\ \ k} w_k \rightarrow I_{ij} + \bar{P} J_{ij}.
\label{eq:DefinitionOfGhostProjection}
\end{align}
Here, $R_{ij}^{\ \ k}$ is the discrete representation of the ensemble of random masks, the integer $k$ indexes the members of this ensemble within the range $[1,N]$, and $w_{k}$ are the non-zero weights such that the above sum approaches the image $I_{ij}$, plus an offset that is equal to the average,
\begin{equation}
\bar{P} = J^{ij} P_{ij} /(nm) = N' \text{E}[w] \text{E}[R].  
\label{eq:Pedestal}
\end{equation}
Above, the matrix $J^{ij}$ is defined to have every element equal to unity and $N'$ is the number of random masks that are prescribed non-zero weights. Furthermore, the units of the pedestal $\bar{P}$ is in terms of the contrast, which we remind the reader is enforced to be unity. In the above and the following equations, we employ the Einstein summation convention in which repeated upper and lower indices are implicitly summed over. Thus, for example, in Eq.~(\ref{eq:DefinitionOfGhostProjection}) the index $k$ is summed over all members of the mask ensemble. Lastly, we note that the weights $w_{k}$ will be proportional to the exposure time for the $k$th random mask. 

We can render Eq.~(\ref{eq:DefinitionOfGhostProjection}) in matrix form by vectorizing our random masks and setting each member equal to the columns of a certain matrix $M$. Further, we can absorb the pedestal $\bar{P}$ into the left-hand side\footnote{Note, we need not necessarily absorb the pedestal $\bar{P}$ into the left-hand side (which, in effect, zero-averages the random masks too). We could instead absorb the pedestal into the image by arbitrarily selecting an acceptable pedestal and numerically optimizing to within this constraint.} by subtracting the average of each column, i.e. 
\begin{equation} \label{eq:Move Pedestal to LHS}
M = [R_{ij1} - \overline{R_{ij1}};R_{ij2}- \overline{R_{ij2}}; \cdots ;R_{ijN}- \overline{R_{ijN}}]. 
\end{equation}
With this, we can now express our ghost projection as
\begin{align}
M \vec{w} \rightarrow \vec{I},
\end{align}
where $\vec{I}$ is the vectorized version of our desired image. With this representation of ghost projection, we can now numerically seek the weights that will achieve
\begin{align}
\text{arg~min} \|M\vec{w} - \vec{I}\|,
\label{eq:OptimisationProblemofGhostProjection1}
\end{align}
subject to the constraint of non-negative weights
\begin{equation}
w_k \geq 0,
\label{eq:OptimisationProblemofGhostProjection2}
\end{equation}
which correspond to physical exposures. A variety of numerical-optimization approaches can be employed to determine the weights $w_k$.  In this investigation, we use non-negative least squares (NNLS) to seek the ghost projection that satisfies Eqs.~(\ref{eq:OptimisationProblemofGhostProjection1}) and (\ref{eq:OptimisationProblemofGhostProjection2}).

Simulations of ghost-projection superpositions, consisting of numerically-generated masks
in different transverse positions, were provided in our previous two papers on ghost projection \cite{paganin2019writing,ceddia2022ghost}.  We refer the reader to these previously-published simulations for an illustration of the application of the theory that has just been outlined, albeit in an artificially-simple context that does not directly utilize experimental data.

\section{Experimental Random Masks}

A fundamental parameter that can be tuned in ghost projection is the set of patterned illuminations available for linear combination to produce the desired ghost-projection image via Eq.~(\ref{eq:DefinitionOfGhostProjection}). Since the objective in this paper is to understand the experimental aspects of ghost projection, although parameter exploration here is by simulation, we employ sets of masks generated from experimental x-ray images. We have opted to utilize commonly available materials that could serve as good candidates for random pattern generation under hard-x-ray illumination, namely (i) a metallic Ni foam slab and (ii) a 120 grit sandpaper sheet.

We have taken large overview (or master mask) images of these random-mask materials and assumed that the ghost projection field-of-view (FOV) is a small subset of this overview. The FOV can be scanned in both horizontal and vertical directions to produce sets consisting of thousands of possible patterned illuminations.  In both of the two cases reported below, images were recorded in the near field, namely for sufficiently short mask-to-detector distances that the morphology of the registered mask image bears a direct resemblance to the projected structure of the mask.

\begin{figure*}[ht!]
     \centering
     \begin{subfigure}{0.3\textwidth}
         \centering
         \includegraphics[width=\textwidth]{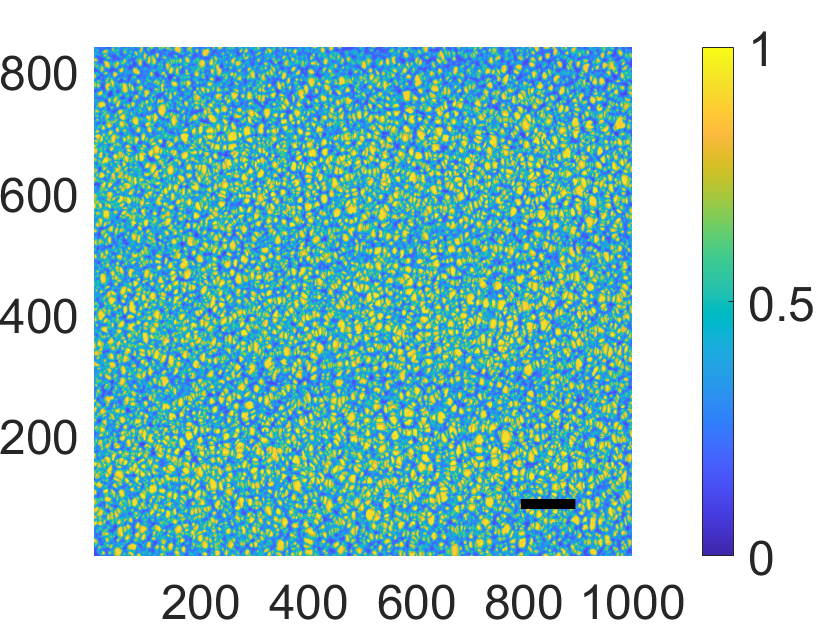}
         \caption{ }
         \label{subfig: RF1}
     \end{subfigure}
     \begin{subfigure}{0.3\textwidth}
         \centering
         \includegraphics[width=\textwidth]{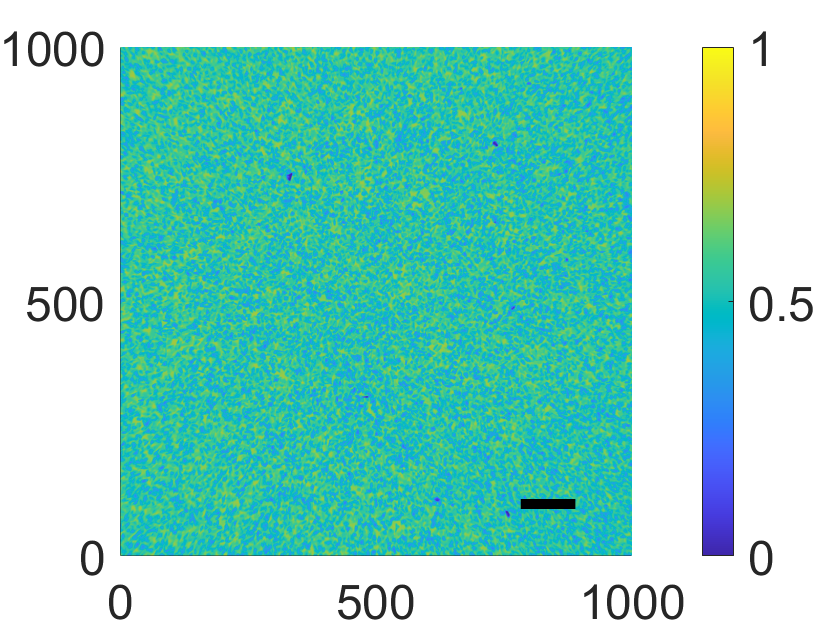}
         \caption{ }
         \label{subfig: SP1}
     \end{subfigure}
     \begin{subfigure}{0.3\textwidth}
         \centering
         \includegraphics[width=\textwidth]{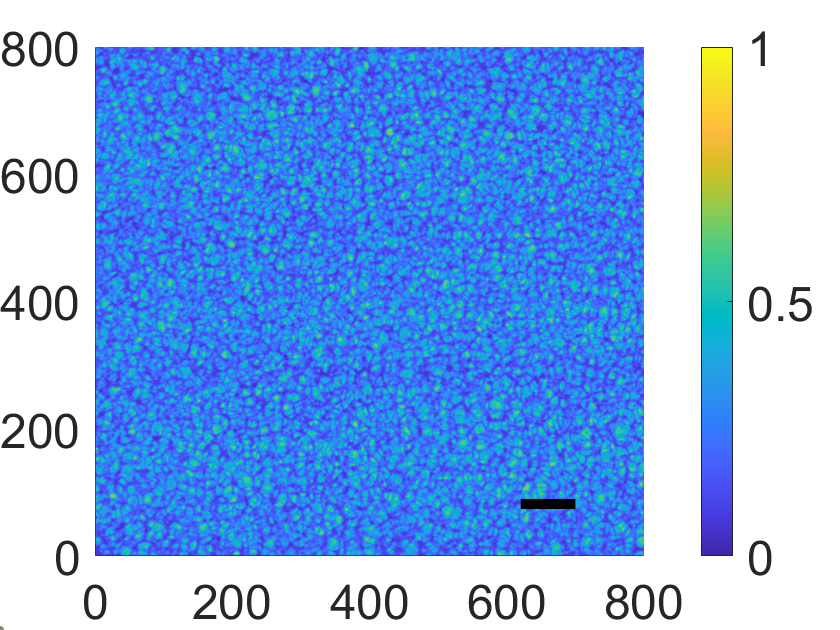}
         \caption{ }
         \label{subfig: CM1}
     \end{subfigure}
     \begin{subfigure}{0.3\textwidth}
         \centering
         \includegraphics[width=\textwidth]{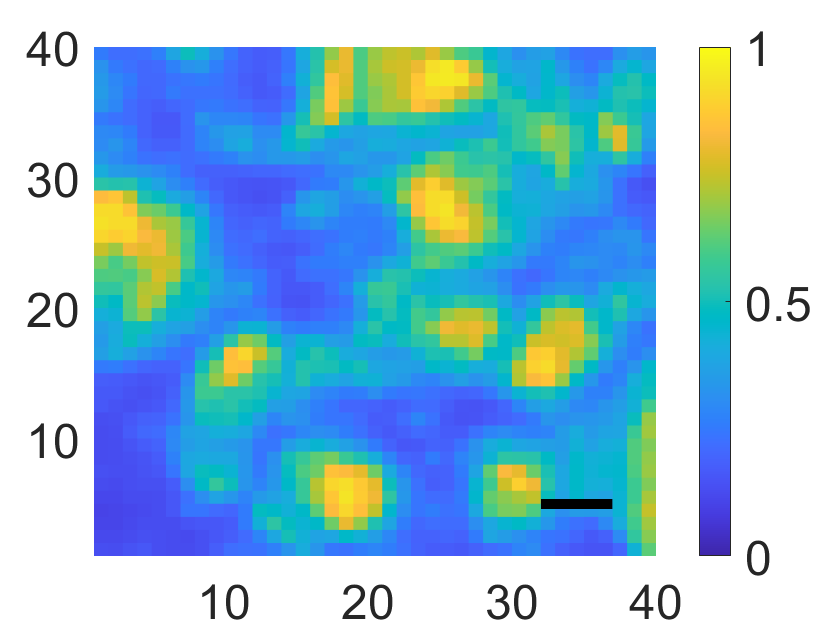}
         \caption{ }
         \label{subfig: RF2}
     \end{subfigure}
     \begin{subfigure}{0.3\textwidth}
         \centering
         \includegraphics[width=\textwidth]{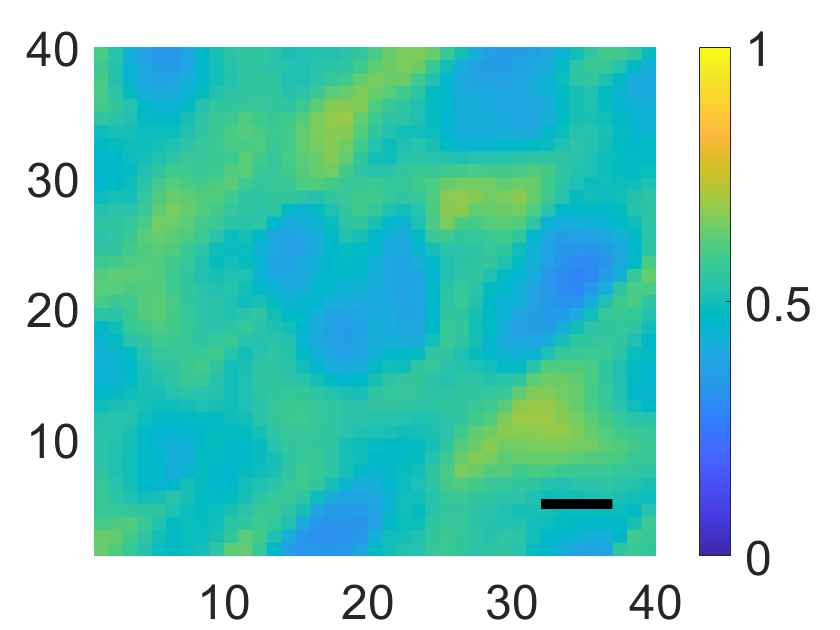}
         \caption{ }
         \label{subfig: SP2}
     \end{subfigure}
     \begin{subfigure}{0.3\textwidth}
         \centering
         \includegraphics[width=\textwidth]{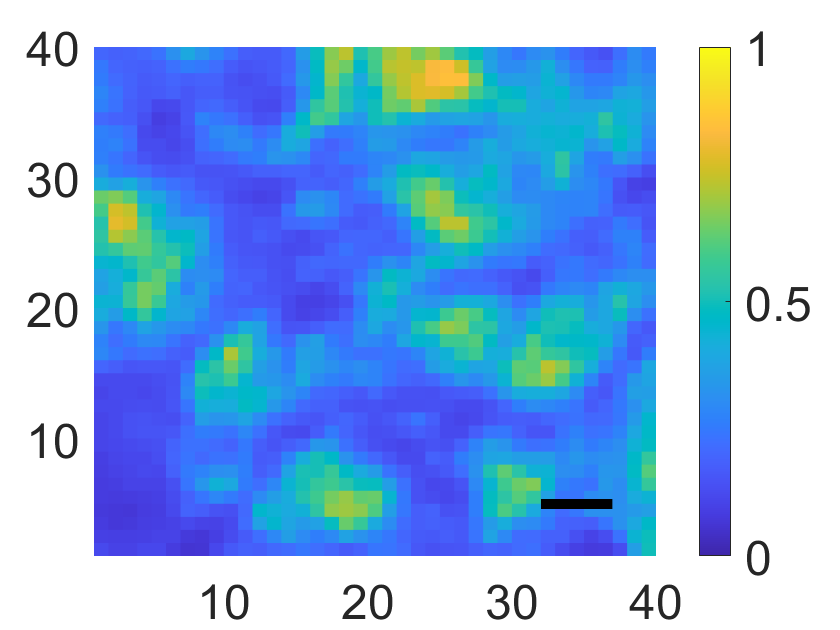}
         \caption{ }
         \label{subfig: CM2}
     \end{subfigure}
     \begin{subfigure}{0.3\textwidth}
         \centering
         \includegraphics[width=\textwidth]{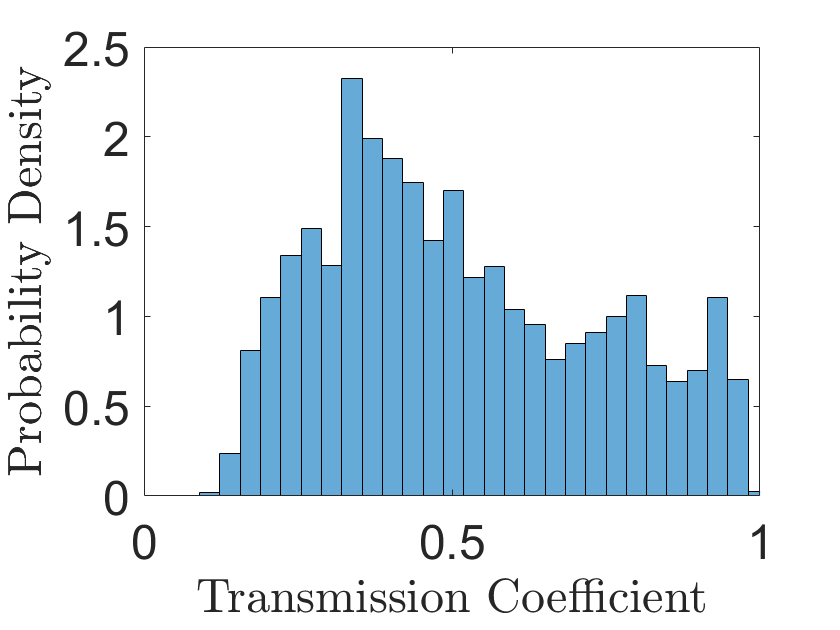}
         \caption{ }
         \label{subfig: RF3}
     \end{subfigure}
     \begin{subfigure}{0.3\textwidth}
         \centering
         \includegraphics[width=\textwidth]{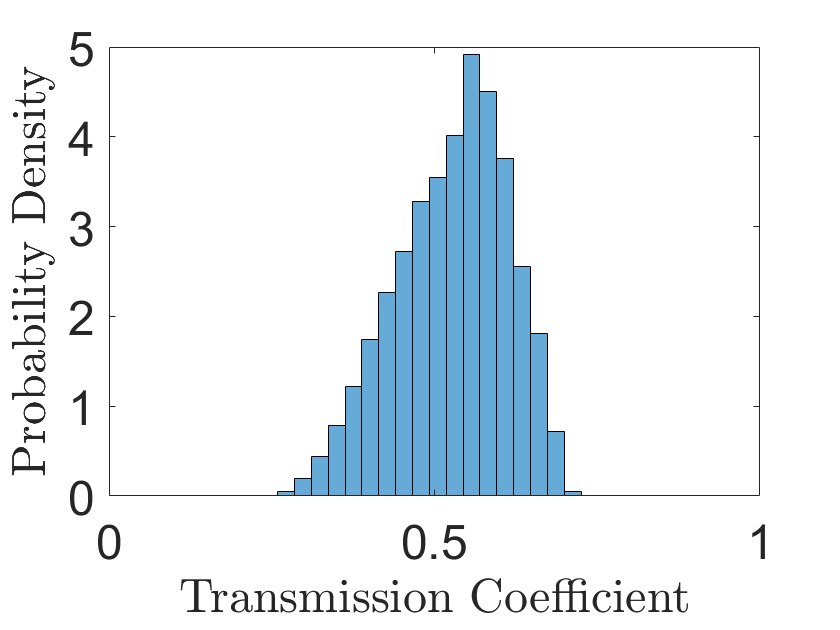}
         \caption{ }
         \label{subfig: SP3}
     \end{subfigure}
     \begin{subfigure}{0.3\textwidth}
         \centering
         \includegraphics[width=\textwidth]{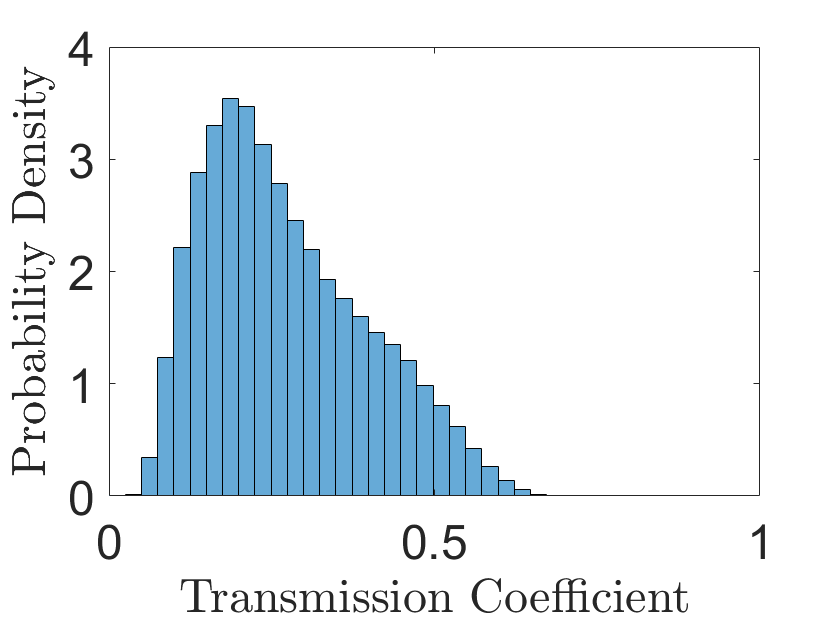}
         \caption{ }
         \label{subfig: CM3}
     \end{subfigure}
     \begin{subfigure}{0.3\textwidth}
         \centering
         \includegraphics[width=\textwidth]{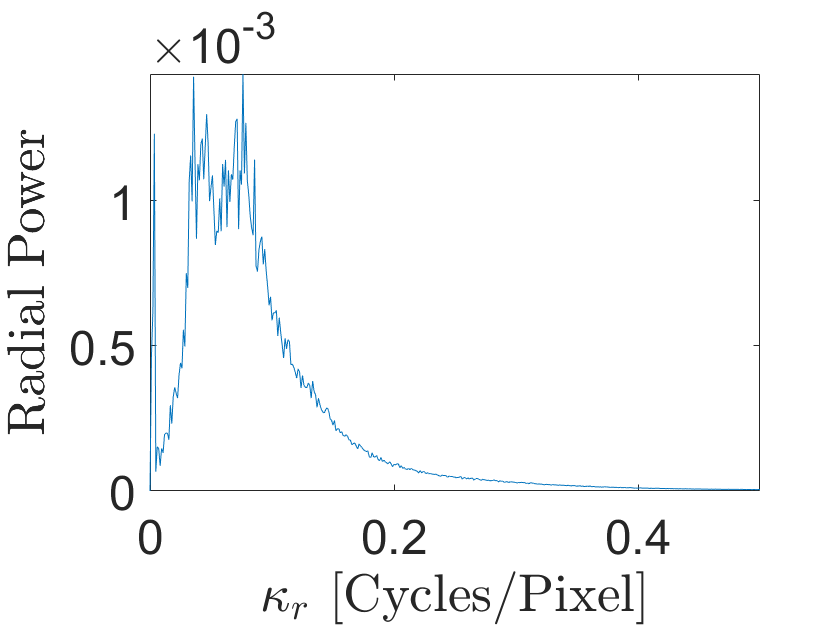}
         \caption{ }
         \label{subfig: RF4}
     \end{subfigure}
     \begin{subfigure}{0.3\textwidth}
         \centering
         \includegraphics[width=\textwidth]{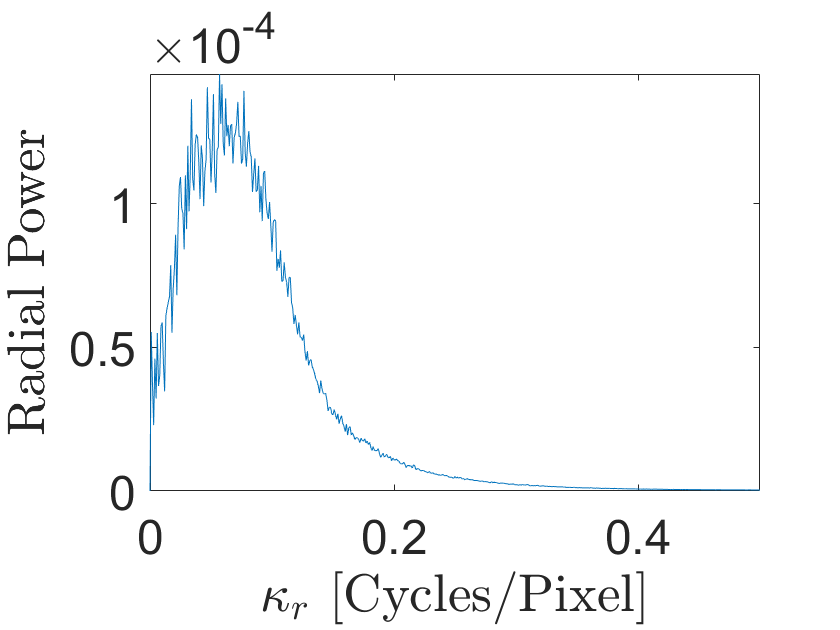}
         \caption{ }
         \label{subfig: SP4}
     \end{subfigure}
     \begin{subfigure}{0.3\textwidth}
         \centering
         \includegraphics[width=\textwidth]{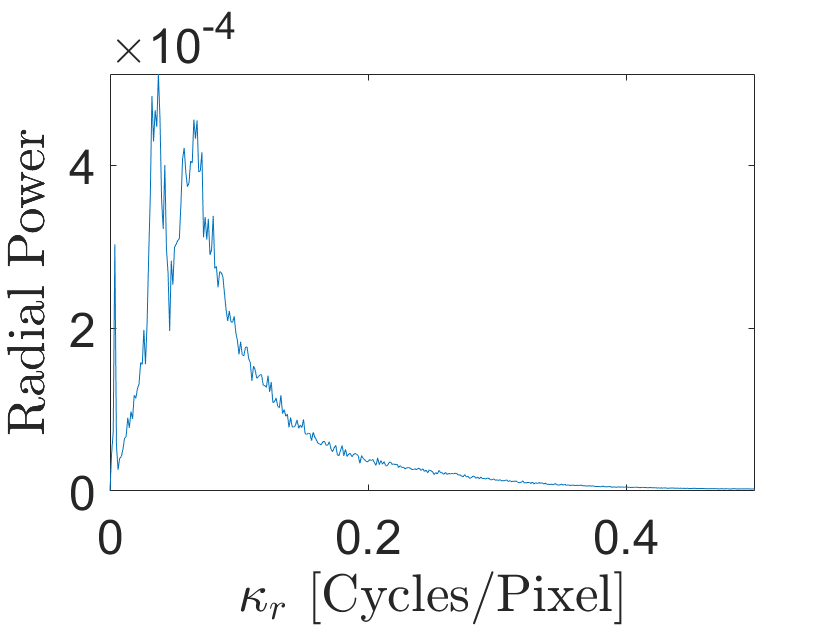}
         \caption{ }
         \label{subfig: CM4}
     \end{subfigure}
        \caption{(a-c) Ni foam, 120 grit sandpaper and consecutive Ni foam and sandpaper master masks. The scale bars are 3mm (30$\mu$m/pixel), 1.248mm (12$\mu$m/pixel) and 2.4mm (30$\mu$m/pixel), respectively. (d-e) $40 \times 40$ field of view (FOV) used to perform ghost projection sampled from the master masks (a-c). Scale bars are 150$\mu$m, 60$\mu$m and 150$\mu$m, respectively. (g-i) Histogram of transmission coefficient of (a-c). (j-l) Angularly integrated 2-dimensional Fourier power spectrum corresponding to panels (a-c), expressed as a function of dimensionless radial spatial frequency $\kappa_r$.}
        \label{fig:Experimental MAsks}
\end{figure*}

The master-mask image of the metallic (Ni) foam is presented in Fig.~\ref{subfig: RF1}. This was recorded at beamline ID19 of the European Synchrotron (ESRF), using a 26keV monochromatic x-ray beam. For further details regarding the setup of the beamline at the time of the experiment, see Weitkamp {\em et al.}~\cite{WeitkampID19}. The detector employed was a FReLoN  (Fast Readout Low Noise) detector developed at ESRF. This is a 14 bit dynamic CCD camera, with a $2\,048 \times 2\,048$ pixel chip and 14$\mu$m pixel pitch. A 0.01s exposure time was used with a 750 mu LuAG:Ce single-crystal scintillator to convert x rays to visible light and optics yielded an effective pixel pitch of 30$\mu$m. An example $40 \times 40$ pixel subset of this master mask used for ghost projection simulation is presented in Fig.~\ref{subfig: RF2}. Figures \ref{subfig: RF3} and \ref{subfig: RF4} present a histogram of the normalized transmission values and the Fourier power spectrum of the Ni foam master mask, respectively. The power spectrum was calculated by taking the two-dimensional Fast Fourier Transform (FFT) \cite{Press2007} of the Ni foam master mask, and then azimuthally averaging the squared modulus of the resulting Fourier-space distribution.

The master-mask image of the 120 grit sandpaper is presented in Fig.~\ref{subfig: SP1}. This was recorded on a custom-built x-ray system at the Australian National University (ANU) CTLab, using a Hamamatsu microfocus x-ray source with a W transmission target. The x rays generated by a 60kV accelerating voltage originate from a 2-3$\mu$m diameter focal point and were filtered by 0.5mm Al. The sandpaper was placed approximately 9mm from the source. A 4343CB Varian a-Si Flat Panel was used to detect the x rays. This has a $3\,040 \times 3\,040$ pixel array with a 0.139mm pixel pitch and uses a CsI scintillator. The detector was placed 315mm from the source and a 32s exposure time was used. The image recorded therefore had an effective pixel pitch of 4$\mu$m. The image presented in  Fig.~\ref{subfig: SP1} has been binned $3 \times 3$. An example $40 \times 40$ pixel subset of this master mask used for ghost projection simulation is presented in Fig.~\ref{subfig: SP2}. Figures \ref{subfig: SP3} and \ref{subfig: SP4} present a histogram of the normalized transmission values and the Fourier power spectrum of the sandpaper master mask, where the power spectrum was calculated in the same fashion as for the Ni foam mask. 

In order to reasonably simulate employing both the Ni foam and sandpaper masks in the same experimental set-up, an imaging-energy correction was applied to the sandpaper transmission values in post processing to make them consistent with the imaging energy used to obtain the Ni foam transmission values. Specifically, this involved raising the transmission values of the sandpaper master mask to the power of 2.04.

To investigate the effect of employing more than one master mask, the metallic foam and sandpaper master masks were placed sequentially to create a consecutive-master-mask. This gives insight into the effect of having multiple length scales present in the resulting random masks. Moreover, the relatively fine and coarse speckles are independently translatable, leading to exponentially more possible unique FOVs. Finally, the attenuation of combined masks should enable sharper, higher-contrast speckles which we hope should reduce the pedestal. For the consecutive-masks case, in order to make the metallic foam speckles and the 120 grit speckles on the same physical scale (originally $30\mu$m/pixel and $4\mu$m/pixel, respectively) the sandpaper master mask was binned from an original size of $3\,040 \times 3\,040$ down by a factor of 7 (i.e. a square of $7 \times 7$, or 49 pixels was averaged into one $28\mu$m pixel). This is as close to the $30\mu$m pixel pitch of the foam image achievable, without resorting to interpolation. This yielded a $434 \times 434$ sandpaper master mask which was subsequently cropped to $400 \times 400$. The $400 \times 400$ sandpaper mask was then tiled on a $800 \times 800$ FOV in Fig.~\ref{subfig: CM1}. A subset of this can be seen in Fig.~\ref{subfig: CM2}. A histogram of the normalized transmission values and power spectrum of the consecutive master mask is presented in Fig.~\ref{subfig: CM3} and \ref{subfig: CM4}, respectively. 

A summary of mask properties including average transmission, variance, peak frequency and weighted average frequency plus-or-minus one standard deviation is presented in Table \ref{tab: mask properties}.  This table also includes entries for delroughness, a quantity that will be introduced in Eq.~(\ref{eq:Delroughness definition}) below.

Free-space propagation effects \cite{Bremmer1952,Snigirev1995,Cloetens1996,Wilkins1996,KleinOpat1976} are inherently accounted for when the projection plane is placed in the same plane as the mask-imaging plane. Such effects may be termed `propagation-based phase contrast', since the local concentration or rarefaction of energy density upon free-space travel converts phase variations into corresponding intensity variations \cite{Paganin2006}.  This same effect has also been denoted as `out-of-focus contrast' \cite{CowleyBook}, and can be thought of as the local focusing or defocusing of radiation or matter waves as they propagate. Propagation-based phase contrast is particularly apt for boosting the fine spatial detail in a speckle mask, since, in the near-field regime where the Fresnel number\footnote{The Fresnel number $N_{\textrm{F}}$ may be defined as $N_{\textrm{F}}={\ell}^2/(\lambda \Delta)$, where (i) $\ell$ is the smallest transverse length scale that is present to a non-negligible degree in the complex wavefield amplitude at the exit surface of the mask, (ii) $\lambda$ is the wavelength or de Broglie wavelength of the radiation or matter waves, respectively, and (iii) $\Delta$ is the propagation distance from the exit surface of the mask to the detection plane.  }\cite{SalehTeichBook} is appreciably greater than unity  and the degree of spatial coherence is sufficiently high, the measured contrast is proportional to the Laplacian of the pre-propagation phase map \cite{Bremmer1952,Paganin2006,Wilkins1996}.  Bearing this phase-contrast effect in mind, it becomes clear that, when the projection plane is longitudinally displaced along the optical axis relative to the imaging plane, this distance should be accounted for when considering the distribution of radiant exposure that the masks will create. For the purposes of this preliminary investigation, the effect of propagation-based phase contrast \cite{Bremmer1952,CowleyBook,Snigirev1995,Cloetens1996,Wilkins1996} for the consecutive masks is ignored. Similar remarks apply to penumbral blurring effects \cite{Delchar1997} associated with a finite source size.

%
%


\begin{table*}[ht!]
\renewcommand{\arraystretch}{1.2}
\begin{tabular}{|c|c|c|c|}
\hline
Mask     & Ni Foam   & 120 Grit Sandpaper   & Consecutive         \\ \hline
E$[R]$   & 0.5091              & 0.5280               & 0.2724              \\ \hline
Var$[R]$    & 0.0489              & 0.0072               & 0.0155              \\ \hline
Delroughness $\mathscr{D}$ & 0.0139        & 0.0022  &  0.0054  \\ \hline
Mode{[}$\kappa_r${]} {[}cyc./pix.{]} & 0.0762              & 0.0570               & 0.0375              \\ \hline
\begin{tabular}[c]{@{}c@{}}E$[\kappa_r] \pm \sqrt{\text{Var}[\kappa_r]}$ \\ {[}cyc./pix.{]}\end{tabular} & $0.0907 \pm 0.0679$ & $0.0853 \pm 0.0610$ & $0.0956 \pm 0.0758$ \\ \hline
\end{tabular}
\caption{Properties of Ni foam, 120 grit sandpaper and consecutive master masks. Note, the definition of delroughness is given in Eq.~(\ref{eq:Delroughness definition}) whereby the values quoted here are taken for the entire masks expressed in Fig.~\ref{subfig: RF1}-\ref{subfig: CM1}. Moreover, these values are only typical of what we might expect from a filtered set of sub-FOVs of these masks, when employed for any particular ghost projection.}
\label{tab: mask properties}
\end{table*}

\vspace*{2em}

We close this section by expanding on the previously-mentioned point that, for both the synchrotron-source and laboratory-source x-ray experiments, the mask measurements were taken in the near-field regime.  This corresponds to the Fresnel number $N_{\textrm{F}}$ being significantly greater than unity.  Moreover, for our masks the projection approximation \cite{Paganin2006} is clearly valid. Such criteria ensure that the morphology of the measured spatially random mask images bear a direct (i.e.~spatially local) resemblance to the projected complex potential of the mask, with the said projection being taken in the direction of the optical axis.  This condition is not necessary, and in principle our method could also be employed using intermediate-field or far-field speckle, namely speckle fields for which the Fresnel number is on the order of unity or much less than unity, respectively. 

\section{Simulations}

Using the experimentally acquired random masks, we numerically optimize ghost projection in a noise-free environment, and then simulate the influence of several noise contributions. These noise contributions are (i)  Poisson noise associated with the finite number of radiation quanta streaming through the ghost-projection system, (ii) exposure-duration noise in the experimental realization of a shutter, and (iii) translational perturbations in the transverse positioning of the random masks.

\vspace*{4em}

\subsection{Experimental noise}

\subsubsection{Poisson noise}
Poisson noise was included in the following manner:
\begin{align}
P_{ij} =  J^{k} \hat{P}( \lambda w_k R_{ijk}  ),
\end{align}
where $\hat{P}$ denotes a quantity that is Poisson distributed, $w_k$ are our numerically derived non-negative weights which correspond to random-mask exposure times, $R_{ijk}$ is a random mask and $\lambda$ is the number of photons per pixel. For this investigation, we employ a value of $\lambda = 10\,000$ photons to contribute to the image contrast. Assuming a perfect or near perfect random mask reconstruction, the dominant contribution to Poisson noise is typically the pedestal which deposits photons without contributing to the contrast of the ghost projection. With this approximation, we can say that the effect of Poisson noise is
\begin{align} \label{eq:Poisson noise}
    J^{k} \hat{P}( \lambda w_k R_{ijk}  ) \rightarrow  \left( \lambda I_{ij} + \lambda \bar{P} \right)  \pm \sqrt{\lambda \bar{P}},
\end{align}
where we have stated this to plus-or-minus one standard deviation. 

\subsubsection{Exposure noise}
Exposure noise, namely the fluctuations in exposure time that will be present in any experimental realization of ghost projection, can be accounted for via
\begin{align}
P_{ij} =  \tilde{P}( w_k ) R_{ij}^{\ \ k},
\end{align}
where $\tilde{P}$ denotes a quantity that is normally distributed such that $\tilde{P}( w_k )$ has an expectation value of the exposure $w_k$ and a corresponding variance $\sigma_w^2$, i.e.~$ \text{E}[\tilde{P}(w_k)] = w_k $, and $\text{Var}[\tilde{P}(w_k)] = \sigma_w^2$, and $R_{ij}^{\ \ k}$ are our random masks. Assuming a perfect or near perfect random mask reconstruction, exposure noise is well described by 
\begin{align}
    \tilde{P}( w_k ) R_{ij}^{\ \ k} \rightarrow \left( I_{ij}  + \bar{P} \right) \pm \sqrt{\sigma_{w}^2 N' \text{Var}[R]},
    \label{eq: analytical exposure noise prediction}
\end{align} 
to one standard deviation, where $N'$ is the number of non-zero weights and Var$[R]$ is the variance of the random mask values. For the purposes of this investigation, we employ an exposure standard deviation of $\sigma_{w} = 1/100$, which is well shy of typical visible-light shutter speeds but reflects the difficulties of controlling high-energy radiation.

\subsubsection{Translational noise}
Translational noise is the error that arises in physically re-aligning the random masks to the positions in which they were imaged. This can be accounted for via
\begin{align}
P_{ij} =  w_k \tilde{R}_{\tilde{i}\tilde{j}}^{\ \ k},
\end{align}
where $w_k$ are the numerically derived exposure times and $\tilde{R}_{\tilde{i}\tilde{j}}^{\ \ k}$ are the interpolated random masks at the perturbed query points which are Gaussian distributed, i.e.~$\tilde{i}\tilde{j} = \tilde{P}( i)\tilde{P}(j )$ has an expectation value of the position $ij$ and a corresponding variance $\sigma_{ij}^2$ in each of the $i$ and $j$ directions. Assuming a perfect or near perfect starting representation of the desired image, the effect of translational noise is well approximated by the first order expression
\begin{align}
    w_k \tilde{R}_{\tilde{i}\tilde{j}}^{\ \ k} \rightarrow \left( I_{ij}  + \bar{P} \right) \pm \sigma_{ij}   w_k \mathscr{D}^k,
    \label{eq: analytical translational noise prediction}
\end{align}
stated to plus-or-minus one standard deviation, where
\begin{align} \label{eq:Delroughness definition}
   \mathscr{D}^k \equiv \sqrt{\text{Var} \left[  \partial_i R_{ij}^{ \ \ k} + \partial_{j} R_{ij}^{ \ \ k} \right]}
\end{align}
is a quantity we call delroughness, so named as roughness may be defined as the standard deviation of a surface profile \cite{NietoVesperinasBook} and delentropy is the entropy of the gradient field \cite{larkin2016reflections}, $\partial_{i}$ is the partial derivative in the $x$-direction index $i$ and similarly for $\partial_{j}$ in the $y$-direction index $j$. Note, the variance stated here is taken over the spatial domain $ij$ and the mask index $k$ is summed over. The derivation of this result is carried through in greater detail in the Appendix.

When numerically perturbing the random masks, a cubic spline was fitted for the purposes of interpolation. To estimate a realistic range for perturbation values, we consulted the quoted specifications for commercially available precision x-y-stages. Kohzu Precision Co.~advertises x-y stages that have repeatability values of $\leq$0.5-0.2$\mu$m for stages that have a motion range of $\pm$30mm. Taking an upper bound value of 0.5$\mu$m, this amounts to a repeatability of 1/60 of a pixel for the Ni foam and 1/24 of a pixel for the 120 grit sandpaper. Being conservative, we will employ a value of $\sigma_{ij} = 1/10$ of a pixel for repeatability of aligning the stage. 

\subsubsection{All contributions of noise}

For ghost projection with Poisson noise in the photon counts, Gaussian noise in the exposure realizations, and Gaussian noise in the position re-alignments, we have
\begin{align}
    P_{ij} = J^{k} \hat{P}( \lambda  \tilde{P}( w_k )  \tilde{R}_{\tilde{i}\tilde{j}k}  ).
\end{align}

\subsubsection{Signal-to-Noise Ratio (SNR)}

We define the signal-to-noise ratio (SNR) of a ghost projection as the root-mean-square (RMS) of the desired image $I_{ij}$ divided by the standard deviation of the projection $\sqrt{\text{Var}[P_{ij}]}$:
\begin{align} \label{eq:SNRdefinition}
     \text{SNR} &=  \sqrt{\frac{1}{nm} J^{ij} \frac{I_{ij}^2}{\text{Var}[P_{ij}]}} = \sqrt{ \frac{\text{E}[I^2] }{\text{Var}[P_{ij}]}}.
\end{align}
Here, $nm$ is the number of pixels of the desired image.  We have used the RMS value since the desired image is zero-centered, i.e.~E$[I]=0$, by construction. Moreover, we have excluded the pedestal as a noise contribution, as for many applications it is not inherently an obstacle.  For example, in a lithographic context one can tune the activation energy of the projection medium or employ additional lithographic substrate.

Note, SNR defined in this fashion will be a function of the resolution of the ghost projection. For example, a ghost projection performed at a resolution of $40 \times 40$ may have a higher SNR compared to the same ghost projection performed at a resolution of $120 \times 120$ for the same FOV. That is, the $40 \times 40$ ghost projection may contain noise at the sub-pixel level that integrates to zero for that pixel. If this is a concern, so long as the masks are sufficiently resolved that they are smoothly and gradually varying between pixels, then the sub-pixel noise should not be problematic.

\subsection{Mask parameters}

\subsubsection{Mask composition}

\begin{figure*}[ht!]
     \centering
     \begin{subfigure}{0.329\textwidth}
         \centering
         \includegraphics[width=\textwidth]{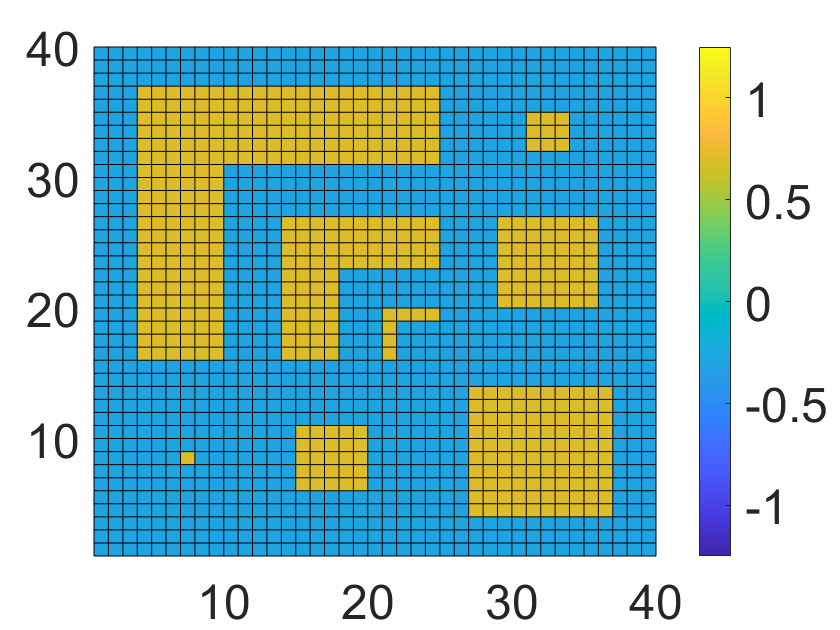}
         \caption{ }
         \label{subfig: GP1}
     \end{subfigure}
     \begin{subfigure}{0.329\textwidth}
         \centering
         \includegraphics[width=\textwidth]{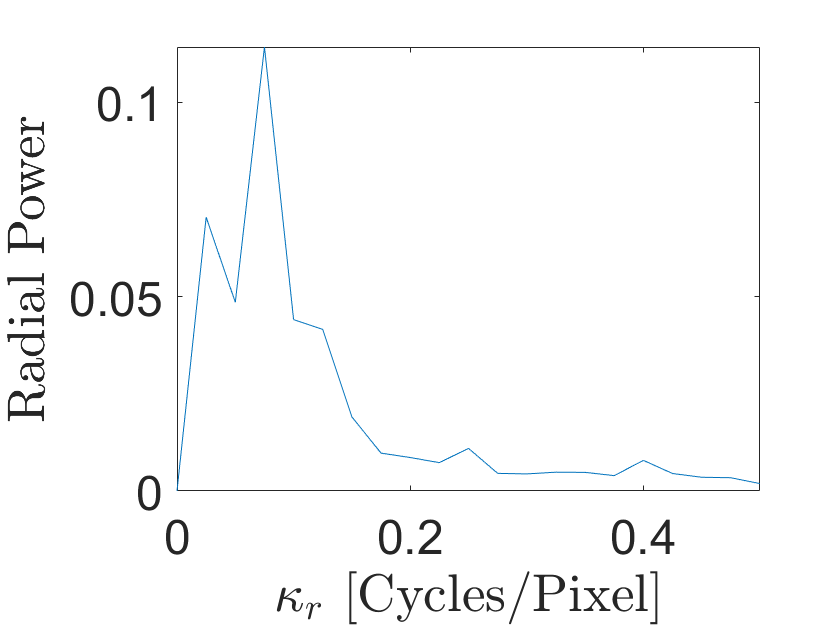}
         \caption{ }
         \label{subfig: GP2}
     \end{subfigure}
     \begin{subfigure}{0.329\textwidth}
         \centering
         \includegraphics[width=\textwidth,clip,trim={0.2cm 0cm 0.2cm 0.15cm}]{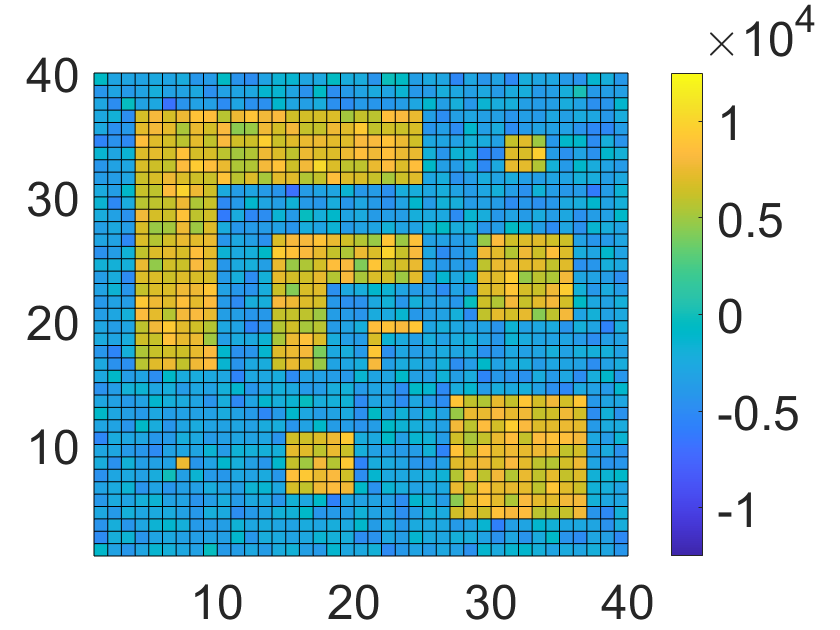}
         \caption{ }
         \label{subfig: GP3}
     \end{subfigure}
     \begin{subfigure}{0.329\textwidth}
         \centering
         \includegraphics[width=\textwidth]{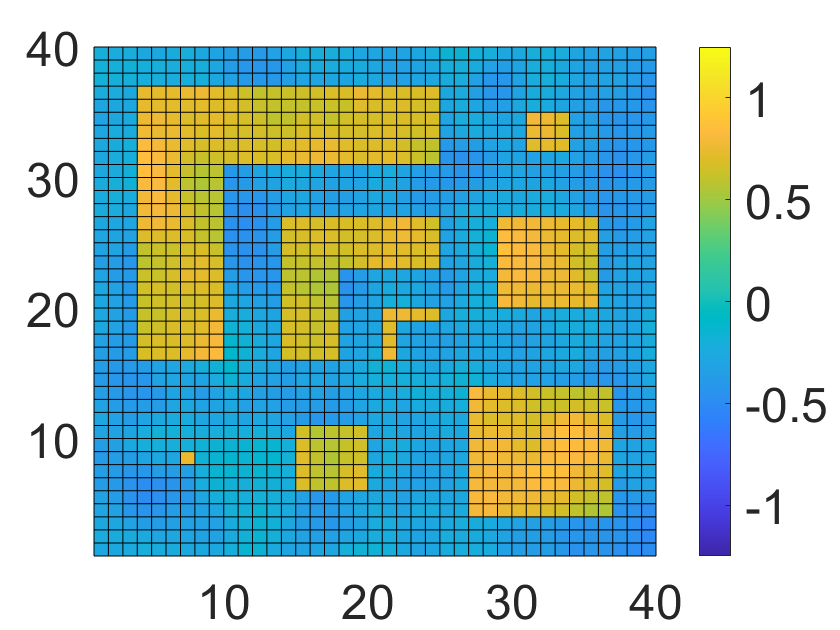}
         \caption{ }
         \label{subfig: GP4}
     \end{subfigure}
     \begin{subfigure}{0.329\textwidth}
         \centering
         \includegraphics[width=\textwidth]{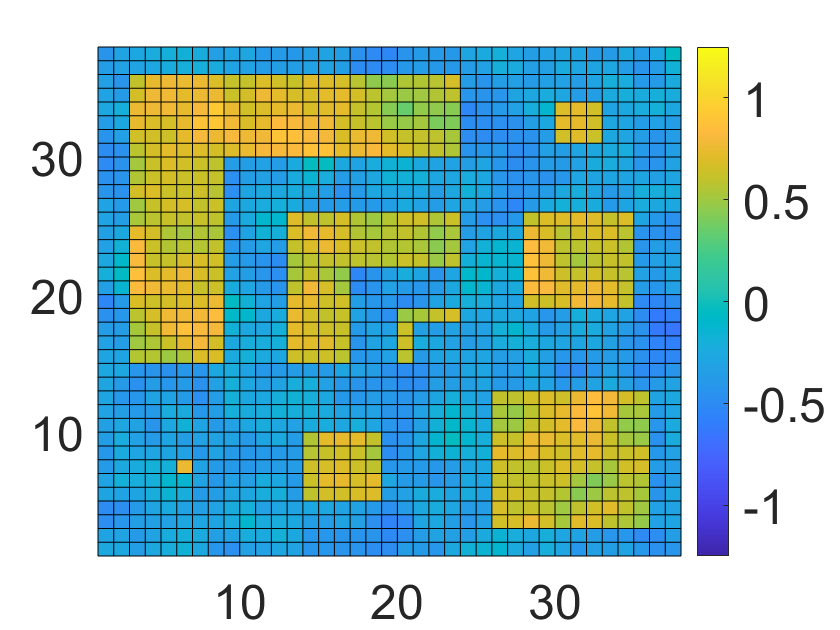}
         \caption{ }
         \label{subfig: GP5}
     \end{subfigure}
     \begin{subfigure}{0.329\textwidth}
         \centering
         \includegraphics[width=\textwidth,clip,trim={0.2cm 0cm 0.2cm 0.15cm}]{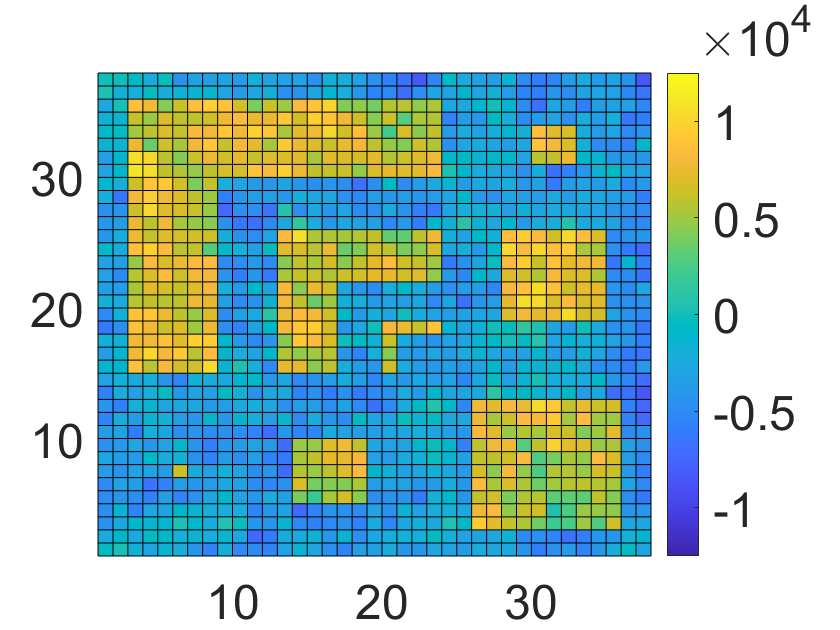}
         \caption{ }
         \label{subfig: GP6}
     \end{subfigure}
        \caption{(a) Non-Negative Least Squares (NNLS) Ni foam ghost projection of a binary image with a contrast of 1, E$[I^2] = 0.2097$, $\text{SNR} = 3.34 \times 10^8$, and pedestal $\bar{P} = 121.7$.  This was constructed using $N = 5nm$ Ni foam masks (recall that $n=40$, $m=40$ is the resolution of our desired image), sampled in increments of 12-by-8 pixels along the $x$ and $y$ axis of the master mask, from which we filtered out $N' = 1\,581$ masks. The physical scale of the projection is 30$\mu$m/pixel. (b) Angularly integrated 2-dimensional Fourier power spectrum of the ghost projection, expressed as a function of dimensionless radial spatial frequency $\kappa_r$. (c) Ghost projection in presence of Poisson noise, with $\lambda = 10\,000$.  (d) Ghost projection in presence of Gaussian exposure noise, with $\sigma_{w} = 1/100$.  (e) Ghost projection in presence of Gaussian translational perturbations, with $\sigma_{ij} = 1/10$. (f) Ghost projection in presence of all noise contributions. The SNR for (c)-(f) is $4.15 \pm 0.07$, $5.62 \pm 0.48$, $4.52 \pm 0.21$, and $2.67 \pm 0.11$, respectively. Note, the translational-perturbations field-of-view is cropped by a border of 1 pixel.}
        \label{fig:six graphs}
\end{figure*}

Consider the $[m \times n] = [40 \times 40]$ binary resolution chart expressed in Fig.~\ref{subfig: GP1}. It is termed a ``binary resolution chart'' as it contains increasingly finely-resolved features that are switched ``on'', situated on a background that is switched ``off''. The contrast of the image is enforced to be unity.  Owing to the zero-centering of the image, namely the constraint that E$[I]=0$, the ``on'' values are approximately 0.7 and the ``off'' values are approximately $-0.3$. 
Moreover, the binary resolution chart expressed in Fig.~\ref{subfig: GP1} is also the NNLS ghost projection made with the Ni foam mask in the absence of experimental noise. Stated differently, we have not included an image of the binary resolution chart and the NNLS reconstruction image separately as, given the SNR of the NNLS reconstruction is $3.34 \times 10^8$, the two images would be indistinguishable by eye. The set of Ni foam sub-FOVs provided to the NNLS optimizer was five times the resolution of the desired image, i.e.~$N = 5nm = 8\,000$. These sub-FOVs were sampled by striding the $40 \times 40$ FOV 12 pixels along the $x$-axis and 8 pixels along the $y$-axis until $5nm$ FOVs were obtained. The NNLS optimizer selected $N' = 1\,581$ mask positions and had a pedestal of 121.7 times the contrast. The corresponding power spectrum of the binary resolution chart is shown in Fig.~\ref{subfig: GP2}. Owing to the sharp edges, we see the presence of high-frequency power-spectrum signal.  Owing to the flat areas of the background and larger features, we also see a significant amount of power-spectrum signal at low-to-moderate spatial frequencies. 

Including the influence of Poisson noise into the near perfect NNLS ghost projection, we obtain Fig.~\ref{subfig: GP3}, which has an SNR of $4.15 \pm 0.07$. The error bounds in the stated SNR were obtained by running the ghost projection simulation with Poisson noise 100 times over, and calculating the corresponding standard deviation in the SNR. Similarly, with suitable adjustments for the type of noise, whenever error bounds are quoted for an SNR, they are obtained in the same fashion as described in the preceding sentence. Examining Fig.~\ref{subfig: GP3}, we can observe that the spatial character of the noise introduced is of the order of the pixel resolution that we are considering. That is, if we attempt to allocate $\lambda = 10\,000$ photons to each pixel to create the desired image contrast, the noise incurred in each pixel is independent of its neighbor and is instead dominated by the size of the ghost projection pedestal.

Compare the spatial character of the noise incurred by Poisson statistics, as shown in Fig.~\ref{subfig: GP3}, to that incurred by exposure noise and translational noise, as shown in Fig.~\ref{subfig: GP4} and \ref{subfig: GP5}. In the latter two cases, the noise instead appears to take on the spatial character of the speckles used to produce the ghost projection. The SNR values in the latter two cases are $5.62 \pm 0.48$ and $4.52 \pm 0.21$, respectively. Including all three types of noise degrades the initial NNLS SNR of $3.34 \times 10^8$ to $2.67 \pm 0.11$. Note, this is not an indication of the best ghost projection of the binary resolution chart possible. Rather, we acknowledge it as an arbitrarily chosen configuration which we regard as a reasonable baseline configuration to gauge our parameter investigation.

To investigate the relative performance of the three types of mask---namely the Ni foam, the sandpaper sheet, and composite mask made by stacking the foam and sandpaper sheets---we simulated a ghost projection of the binary resolution chart made with $N=5nm$ FOVs each. The sandpaper was stepped in 10-by-10 pixel increments between FOVs and for the consecutive mask case, the Ni foam was stepped in 12-by-8 pixel increments as before while the sandpaper, having been binned down relative to the previous case to match the resolution of the Ni foam, was stepped in 4-by-4 pixel increments. The results of the ghost projections performed with the three types of mask are shown in Table \ref{tab: mask comparison}. Looking at the column corresponding to the 120 grit sandpaper, the experimental-noise-free NNLS reconstruction SNR is similar to that of the Ni foam and in each case, a similar number of FOVs were filtered from the $N = 5nm$ set. Where they differ is in the size of the pedestal, which we might have expected in a qualitative sense based on the increased average transmission and lower proportion of high-frequency power present in the sandpaper mask as compared to the Ni foam mask: see Table \ref{tab: mask properties}. Owing to this increased pedestal in the NNLS sandpaper ghost projection, we observe a greater erosion of the noise-free SNR due to Poisson noise than in the Ni foam case. Similarly, we also see a greater erosion of the noise-free SNR due to translational noise in the sandpaper case as compared to the Ni foam case due to the relatively large weights $w_k$ employed. That is, despite the sandpaper having a smaller typical delroughness than the Ni foam (see Table \ref{tab: mask properties}), the larger reconstruction weights dominate the amount of noise we expect from translational noise (see Eq.~(\ref{eq: analytical translational noise prediction})). When comparing the robustness of the two ghost projections in the presence of exposure noise, however, we can see that the sandpaper outperforms the Ni foam. We might have expected this based upon the analytical prediction expressed in Eq.~(\ref{eq: analytical exposure noise prediction}), which states that exposure noise is proportional to $\sqrt{\text{Var}[R]}$, and based on Table \ref{tab: mask properties}, we can see that the variance of the sandpaper mask transmission values is approximately an order of magnitude lower than that of the Ni foam. 

The previous paragraph illuminates a series of trade-offs that a prospective ghost projection realization should consider. On one hand, we want to minimize the following four quantities: (i) the average transmission value E$[R]$, (ii) the transmission variance Var$[R]$, (iii) the delroughness $\mathscr{D}$, and (iv) the number of weights $N'$ (where the number of weights is typically of the order of the resolution we are considering). On the other hand, we also want sufficiently sharp features present in the mask to match the sharpness of features present in the desired projection and we want these to be sufficiently well resolved at the resolution we are considering. The relative importance of which aspects of the mask properties to focus on will depend on the experimental tolerance parameters $\sigma_w$ and $\sigma_{ij}$ as well as the desired resolution. For example, exposure noise may be the dominant contribution to the overall noise and hence, the variance of the mask transmission values should be the focus of effort invested into improving the realized ghost projection. 

For the case of consecutive masks, we can once again see that the experimental-noise-free NNLS SNR is similar in order of magnitude to the previous cases of Ni foam and sandpaper employed individually.  We can also see that the consecutive-masks case filters out a similar number of FOVs $N'$ to the Ni foam and sandpaper cases. In terms of the ghost-projection pedestal, this case has the smallest pedestal which leads us to expect it will be the most robust to noise inclusions. Once the effects of experimental noise were included, this case was seen to outperform the previous two in all respects bar one exception. The combination creates FOVs that meet our desire to lower the average transmission value, increase the relative proportion of high-frequency power and it reduces the variance of the transmission values with respect to the Ni foam mask. Note, the variance of transmission values for the consecutive masks is actually increased with respect to the sandpaper mask and hence performs worse under exposure noise -- this is the exception previously mentioned, regarding the consecutive-masks case outperforming the single-mask case in all respects, for the particular experimental masks studied here. 

\begin{table*}[ht!]
\centering
\renewcommand{\arraystretch}{1.2}
\begin{tabular}{|c|c|c|c|} 
\hline
Mask                 & Ni Foam            & 120 Grit Sandpaper & Consecutive        \\ \hline
NNLS SNR             & $3.34 \times 10^8$ & $3.14 \times 10^8$ & $4.02 \times 10^8$ \\ \hline
$\bar{P}$            & $121.7$            & $417.1$            & $85.0$             \\ \hline
$N'$                 & $1581$             & $1587$             & $1582$             \\ \hline
SNR w/ Poisson       & $4.15 \pm 0.07$    & $2.24 \pm 0.04$    & $4.98 \pm 0.08$    \\ \hline
SNR w/ Exposure      & $5.62 \pm 0.48$    & $13.69 \pm 0.97$   & $9.58 \pm 0.08$    \\ \hline
SNR w/ Translational & $4.52 \pm 0.21$    & $3.26 \pm 0.18$    & $5.38 \pm 0.24$    \\ \hline
SNR w/ All           & $2.67 \pm 0.11$    & $1.83 \pm 0.04$    & $3.39 \pm 0.11$    \\ \hline
\end{tabular}
\caption{Ghost projections with different master masks, $N = 5nm$, strided 12-by-8 for the Ni foam, 10-by-10 for the 120 grit sandpaper, and 12-by-8 and 4-by-4 for the Ni foam and sandpaper when used simultaneously. Poisson noise, exposure noise and translational noise were for the conditions $\lambda = 10\,000$, $\sigma_w = 1/100$ and $\sigma_{ij} = 1/10$, respectively.}
\label{tab: mask comparison}
\end{table*}

\subsubsection{Number of masks and mask striding} \label{subsubsec:No of masks and striding}

Here we investigate the following experimental parameters: (i) the number of sub-FOVs or masks sampled from the master mask, and (ii) the striding employed between these masks. The upper limit of these parameters is to stride at 1-pixel increments and to include as many sub-FOVs as possible. This would give the numerical optimizer all the information needed to find the optimal reconstruction achievable with a given master mask. Depending on the size of the master mask and the number of sub-FOVs, however, this may create such a large set of masks that optimizing the reconstruction becomes exceedingly computationally demanding and may even prove infeasible. This prompted an examination of the effect of the number of masks, together with a consideration of the influence of different striding protocols, for the same master mask (Ni foam) and the same desired projected-image (the binary resolution chart).

Table \ref{tab:number of masks and striding} gives the results of our numerical investigation. The main trend is that the greater the number of masks made available to the numerical optimizer, the more robust the reconstruction is to the presence of noise. That is, using $N=2nm$ we obtained a noise-free SNR of 30.34 and a final SNR of $0.75 \pm 0.03$ once Poisson, exposure and translational noise was included. This increased to a noise-free SNR of $1.67\times 10^7$ with $N = 100nm$ and had a final SNR of $3.30 \pm 0.14$. One exception is the $N=2nm$ case, strode in a 1-by-1 fashion. Here we observe a reconstruction which employs very few of the masks available (625 of the 3\,200) and uses relatively low weights to achieve a very modest experimental-noise-free reconstruction with an SNR of 3.24. However, this modest starting SNR proves to be rather robust to experimental noise, and out-performs many of the cases which have a much high starting SNR. Evidently, a higher noise-free SNR does not necessarily result in a higher noise-included SNR. Our results suggest that the numerical optimizer employed to create the reconstruction is sub-optimal at finding reconstructions that are robust to noise inclusions.  Improvements in this could be the focus of future work. In particular, NNLS simply optimizes the starting SNR whereas an optimizer that accounts for the inclusion of noise holds the potential for significant gains in final SNR values. 

In the case of classical ghost imaging, the natural spatial resolution is that of the size of the speckles present in the random masks \cite{ferri2010differential,PellicciaIUCrJ}. Similarly, it was assumed in our preliminary work on ghost projection \cite{ceddia2022ghost} that the speckle size would form a natural resolution. This is accurate for ghost projection formed by weighting or filtering the random mask set according to a metric proportional to the correlation between the random masks and the desired image. In this investigation, we have found that the resolution of numerically optimized ghost projection is limited by the highest spatial frequencies present to a non-negligible degree in the mask power spectrum, such as is the case with computational ghost imaging \cite{PellicciaIUCrJ}. In this context, we note that the power spectrum of typical experimental speckle masks will extend to radial spatial frequencies beyond that which corresponds to the peak power-spectrum value.  This last comment may be particularly relevant for long-tailed power spectra, for example those with Lorentzian rather than Gaussian form. We speculate that numerical optimization is able to achieve this sub-speckle resolution by using constructive and destructive ``interference''\footnote{Here, the term ``interference'' refers to both additive and subtractive effects associated with superposing real-valued functions (at the level of net ghost-projection radiant exposure), rather than the more usual meaning of additive and subtractive effects in resultant intensity due to the superposition of complex-valued functions.  Our usage of the term is consistent, for example, with the concept of intensity interferometry in the context of the Hanbury Brown-Twiss effect \cite{HanburyBrownBook,mandel1995optical}. } of the speckles, at the cost of an increased pedestal. One approach to test this speculation would be to compare two ghost projections made from the same master mask, assuming both sets of sub-FOVs span the matrix space defined by the image resolution. The first set is comprised of entirely unique, non-overlapping FOVs and the second set has FOVs that are strode systematically such that there is significant overlap between FOVs. The conclusion of this test thus hinges on whether the systematically strode masks generate a smaller ghost projection pedestal and better SNR than the non-overlapping FOV case, implying masks with scanned speckles enable desirable interference. Unfortunately, given the size of the experimentally obtained masks employed here, we cannot sample a sufficiently large number of unique, non-overlapping FOVs to span the matrix space defined by the image resolution and test this.  The closest we can do is to explore a reduced range of striding protocols with the same master mask, all of which still have overlap with other FOVs. In Table \ref{tab:number of masks and striding}, for the $N=5nm$ case, we can observe a small improvement in final SNR with increased striding from 1-by-1 pixels to 12-by-8 pixels, although this does not appear to be indicative of a general trend. 
Seemingly, so long as the set of sub-FOVs adequately spans the matrix space defined by the image resolution, the fashion in which they are sampled is mostly immaterial.

An alternative test for constructive and destructive interference leading to sub-speckle resolution in numerically optimized ghost projection is to examine the distribution of sub-FOVs selected as a function of striding, or transverse position in the case that adjacent FOVs are available. If several masks are sampled systematically within a relatively small neighborhood, this implies that the sub-speckle resolution could indeed be created by the interference of ``scanned'' speckles. If the mask distribution does not display any evidence of being systematically sampled, however, then this would serve as evidence against our hypothesis of interference. The spatial-distribution of sub-FOVs naturally arises in Sec.~\ref{Old AppB} where a traveling salesperson perspective is adopted, and these weights are compared to selecting sub-FOVs uniformly at random there (see Figs.~\ref{Fig:WeightsDistribution} and \ref{fig:TSP}).

\begin{table*}[ht!]
\centering
\renewcommand{\arraystretch}{1.2}
\begin{tabular}{|c|cc|ccc|c|c|c|c|}
\hline
$N$                                                               & \multicolumn{2}{c|}{$2nm$}                             & \multicolumn{3}{c|}{$5nm$}                                                                             & $10nm$             & $20nm$             & $50nm$             & $100nm$            \\ \hline
Striding                                                        & \multicolumn{1}{c|}{1-by-1}          & 6-by-8          & \multicolumn{1}{c|}{1-by-1}             & \multicolumn{1}{c|}{6-by-8}             & 12-by-8            & 1-by-1             & 1-by-1             & 1-by-1             & 1-by-1             \\ \hline
NNLS SNR                                                        & \multicolumn{1}{c|}{3.24}            & 30.34           & \multicolumn{1}{c|}{$4.43 \times 10^8$} & \multicolumn{1}{c|}{$4.27 \times 10^8$} & $3.34 \times 10^8$ & $1.49 \times 10^8$ & $9.07 \times 10^7$ & $2.97 \times 10^7$ & $1.67 \times 10^7$ \\ \hline
$\bar{P}$                                                       & \multicolumn{1}{c|}{31.9}           & 721.8          & \multicolumn{1}{c|}{131.5}              & \multicolumn{1}{c|}{125.4}             & 121.7              & 94.5              & 86.3              & 78.5              & 77.5              \\ \hline
$N'$                                                            & \multicolumn{1}{c|}{625}             & 1581            & \multicolumn{1}{c|}{1582}               & \multicolumn{1}{c|}{1580}               & 1581               & 1565               & 1547               & 1515               & 1491               \\ \hline
\begin{tabular}[c]{@{}c@{}}SNR w/\\ Poisson\end{tabular}        & \multicolumn{1}{c|}{$3.00 \pm 0.03$} & $1.70 \pm 0.03$ & \multicolumn{1}{c|}{$4.01 \pm 0.07$}    & \multicolumn{1}{c|}{$4.09 \pm 0.07$}    & $4.15 \pm 0.07$    & $4.73 \pm 0.08$    & $4.93 \pm 0.09$    & $5.16 \pm 0.10$    & $5.20 \pm 0.09$    \\ \hline
\begin{tabular}[c]{@{}c@{}}SNR w/ \\ Exposure\end{tabular}      & \multicolumn{1}{c|}{$3.03 \pm 0.04$} & $5.26 \pm 0.43$ & \multicolumn{1}{c|}{$5.67 \pm 0.56$}    & \multicolumn{1}{c|}{$5.49 \pm 0.43$}    & $5.62 \pm 0.48$    & $5.71 \pm 0.53$    & $5.87 \pm 0.47$    & $5.89 \pm 0.49$    & $5.91 \pm 0.52$    \\ \hline
\begin{tabular}[c]{@{}c@{}}SNR w/ \\ Translational\end{tabular} & \multicolumn{1}{c|}{$3.05 \pm 0.02$} & $0.84 \pm 0.04$ & \multicolumn{1}{c|}{$4.22 \pm 0.27$}    & \multicolumn{1}{c|}{$4.34 \pm 0.24$}    & $4.52 \pm 0.21$    & $5.54 \pm 0.29$    & $5.96 \pm 0.32$    & $6.39 \pm 0.32$    & $6.29 \pm 0.32$    \\ \hline
\begin{tabular}[c]{@{}c@{}}SNR w/ \\ All\end{tabular}           & \multicolumn{1}{c|}{$2.71 \pm 0.04$} & $0.75 \pm 0.03$ & \multicolumn{1}{c|}{$2.60 \pm 0.12$}    & \multicolumn{1}{c|}{$2.62 \pm 0.10$}    & $2.67 \pm 0.11$    & $3.05 \pm 0.13$    & $3.18 \pm 0.15$    & $3.29 \pm 0.14$    & $3.30 \pm 0.14$    \\ \hline
\end{tabular}
\caption{Investigation into (i) the total number of masks sampled from the master mask using the Ni foam master mask and (ii) the overlap between mask FOVs when performing a ghost projection of the $[m,n] = 40 \times 40$ binary resolution chart. Poisson, exposure and translational noise inclusions were for the conditions $\lambda = 10\,000$, $\sigma_w = 1/100$ and $\sigma_{ij} = 1/10$, respectively.}
\label{tab:number of masks and striding}
\end{table*}

\subsubsection{Combining consecutive masks}

As previously stated at the opening of Sec.~\ref{subsubsec:No of masks and striding}, the maximal protocol for sampling sub-FOVs from the master mask (or masks) is to use all possible configurations, scanned at 1-by-1 pixel increments. In the single-mask case, we noted that this maximal protocol has the potential to create a set of masks so large in number that the computational requirements to perform the numerical optimization may become infeasible.  Moreover, as already noted, the number of available masks grows exponentially, in passing from the case of a single mask to two consecutive master masks. Evidently, some sensible protocol to sample a subset of all possible configurations should be employed. The three protocols explored here are (i) systematically stepping the Ni foam and sandpaper master masks, (ii) systematically stepping the Ni foam master mask while cycling through unique FOVs of the sandpaper master mask, and (iii) randomly selecting FOVs from both the Ni foam and sandpaper master masks. Using these three protocols, we sampled $N=5nm$ masks in order to ghost project the binary resolution chart in Fig.~\ref{subfig: GP1}.

When constructing the systematically sampled consecutive mask, since the Ni foam and sandpaper master masks are of different transverse spatial extents, we stepped each along according to their own scheme. That is, we stepped the Ni foam along in the same 12-by-8 pixel fashion that has been our benchmark, and for the 120 grit sandpaper, we stepped this along in a 4-by-4 pixel fashion until $N = 5nm$ masks had been sampled. Recalling the result from Table \ref{tab: mask comparison}, we saw that the addition of having the sandpaper in conjunction with the Ni foam improved the NNLS reconstruction SNR, decreased the pedestal and almost across the board proved to be more robust to noise inclusions in comparison to just the Ni foam, or sandpaper alone.

When constructing the uniquely sampled consecutive mask, we sampled the Ni foam in the 12-by-8 increments as before but we cut the $400 \times 400$ sandpaper mask into 100 $40 \times 40$ unique FOVs and cycled through these in conjunction with the Ni foam mask FOVs. This approach proved to be a relatively poor performer with the lowest NNLS reconstruction compared to the systematically and randomly constructed consecutive masks. Where this particularly struggled was the sharp edges of the binary resolution chart. While this approach did have similar high-frequency-power compared to the other sampling protocols, we hypothesize that this difficulty with the sharp edges is because of the fashion in which the ``uniquely'' sampled masks were constructed. Since the same 100 FOVs from the sandpaper mask were recycled 80 times over, we conjecture that this inhibited the ability of the set to adequately span the image resolution space -- i.e.~contrary to our intention of constructing the most ``unique'' set, they were less unique than the randomly sampled or even systematically sampled set.

The randomly sampled consecutive mask was constructed by randomly picking a $40 \times 40$ FOV from the Ni foam and combining it with a randomly selected $40 \times 40$ FOV from the 120 grit sandpaper. Of the three protocols for constructing consecutive masks, this performed the best in terms of NNLS SNR, minimal pedestal and noise robustness. Seemingly, similar to our previous investigation into striding, so long as the constructed set of masks spans the desired image resolution space, it does not matter if the masks are sampled in a systematic or random way, so long as they are not inadvertently pathological and violate the condition of spanning the image resolution space.

\begin{table*}[ht!]
\centering
\renewcommand{\arraystretch}{1.2}
\begin{tabular}{|c|c|c|c|}
\hline
\begin{tabular}[c]{@{}c@{}}Combining \\ Consecutive Masks\end{tabular} & Systematically     & Uniquely        & Randomly           \\ \hline
NNLS SNR                                                               & $4.02 \times 10^8$ & 3.22            & $6.03 \times 10^8$ \\ \hline
$\bar{P}$                                                              & 85.0               & 19.1            & 82.4               \\ \hline
$N'$                                                                   & 1582               & 636             & 1582               \\ \hline
SNR w/ Poisson                                                         & $4.98 \pm 0.08$    & $3.07 \pm 0.02$ & $5.06 \pm 0.09$    \\ \hline
SNR w/ Exposure                                                        & $9.58 \pm 0.08$    & $3.14 \pm 0.01$ & $9.80 \pm 0.71$    \\ \hline
SNR w/ Translational                                                   & $5.38 \pm 0.24$    & $3.08 \pm 0.01$ & $5.46 \pm 0.21$    \\ \hline
SNR w/ All                                                             & $3.39 \pm 0.11$    & $2.90 \pm 0.03$ & $3.46 \pm 0.12$    \\ \hline
\end{tabular}
\caption{Comparison of different ways of combining the consecutive masks, with $N= 5 nm$ for the binary resolution chart. The second column corresponds to striding the Ni foam mask along 12-by-8 and 120 grit sandpaper in 4-by-4 increments. The third column corresponds to striding the Ni foam the same as before, however, each of these FOVs is paired with 1 of 100 unique FOVs sampled from the 120 grit sandpaper. The final column corresponds to randomly picking a FOV from each of the Ni foam and 120 grit sandpaper and combining them, until $5nm$ combinations of masks is obtained. Noise contributions from Poisson, exposure and translational are for the conditions $\lambda = 10\,000$, $\sigma_w = 1/100$ and $\sigma_{ij} = 1/10$, respectively.}
\label{tab:Consecutive Masks}
\end{table*}

\subsection{Image parameters}

In this section, we explore parameters associated with the desired image that we wish to ghost project. In loose terms, we perform this exploration by varying the ``amount'' and ``distribution'' of the image. We quantify the ``amount'' of image via the image norm $\sqrt{\text{E}[I^2]}$. Note, what we call image norm is actually an area normalized image norm $|I_{ij}|/|J_{ij}|$, but for the sake of brevity and clarity, we simply refer to this as image norm. We quantify the ``distribution'' of an image via its angularly integrated 2-dimensional Fourier power spectrum, henceforth referred to as an image power spectrum.

\begin{figure*}[ht!]
     \centering
     \begin{subfigure}{0.329\textwidth}
         \centering
         \includegraphics[width=\textwidth]{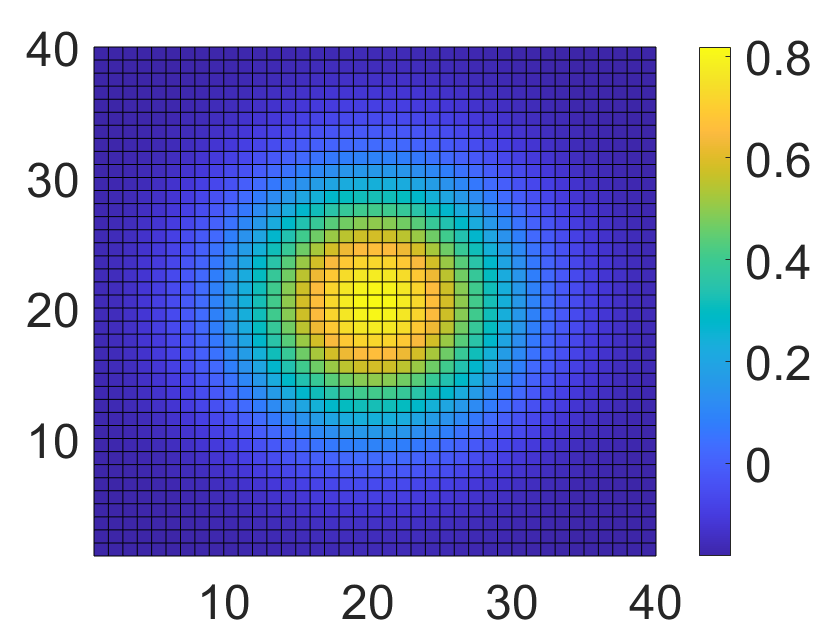}
         \caption{ }
         \label{subfig:GP1 test}
     \end{subfigure}
     \begin{subfigure}{0.329\textwidth}
         \centering
         \includegraphics[width=\textwidth]{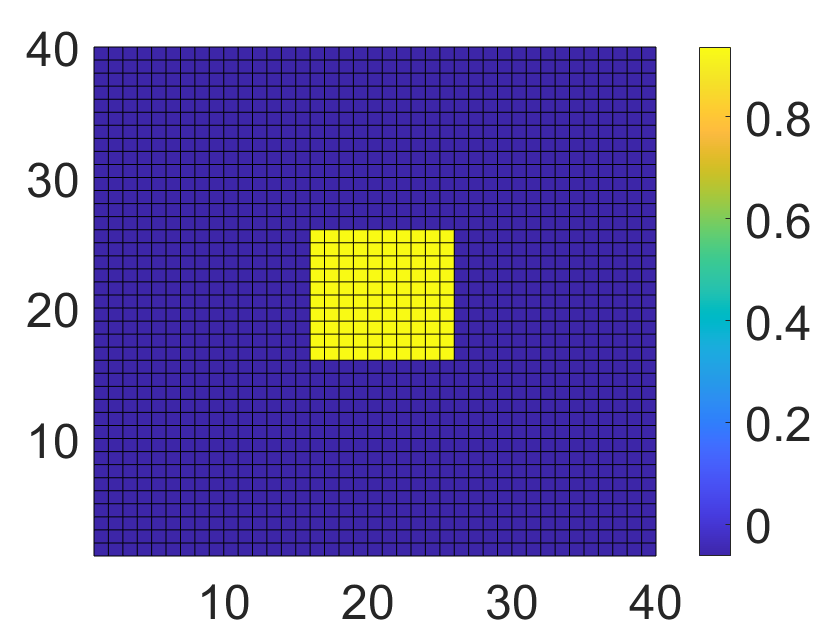}
         \caption{ }
         \label{subfig:GP2 test}
     \end{subfigure}
     \begin{subfigure}{0.329\textwidth}
         \centering
         \includegraphics[width=\textwidth]{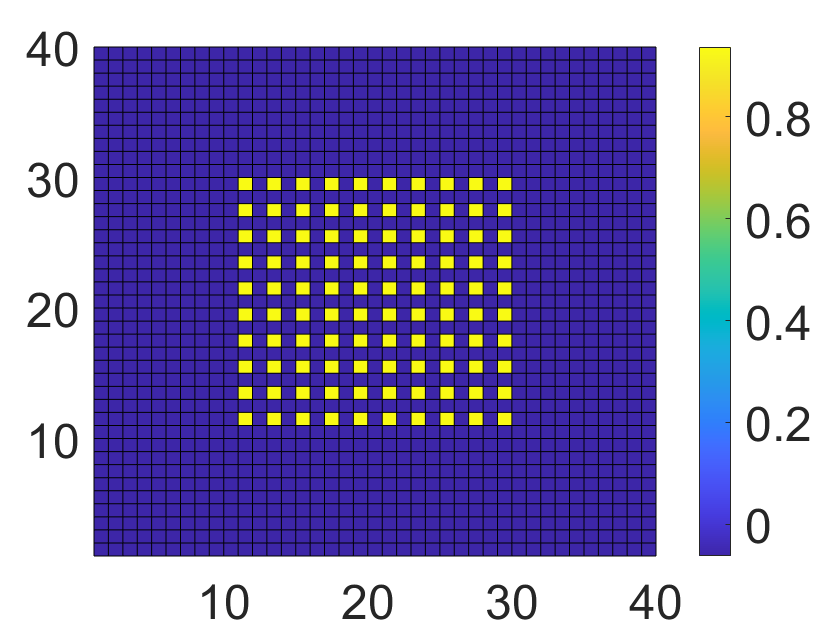}
         \caption{ }
         \label{subfig:GP3 test}
     \end{subfigure}
     \begin{subfigure}{0.329\textwidth}
         \centering
         \includegraphics[width=\textwidth]{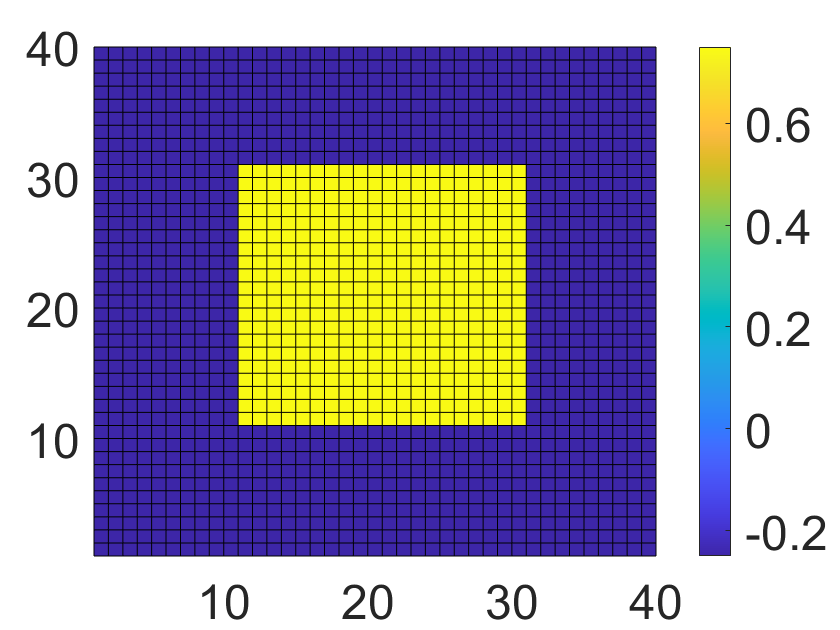}
         \caption{ }
         \label{subfig:GP4 test}
     \end{subfigure}
     \begin{subfigure}{0.329\textwidth}
         \centering
         \includegraphics[width=\textwidth]{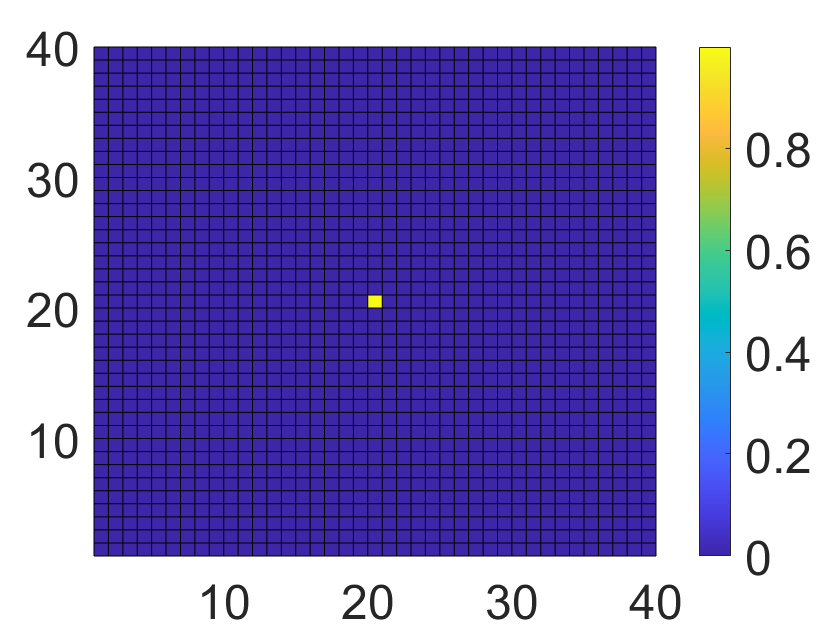}
         \caption{ }
         \label{subfig:GP5 test}
     \end{subfigure}
     \begin{subfigure}{0.329\textwidth}
         \centering
         \includegraphics[width=\textwidth]{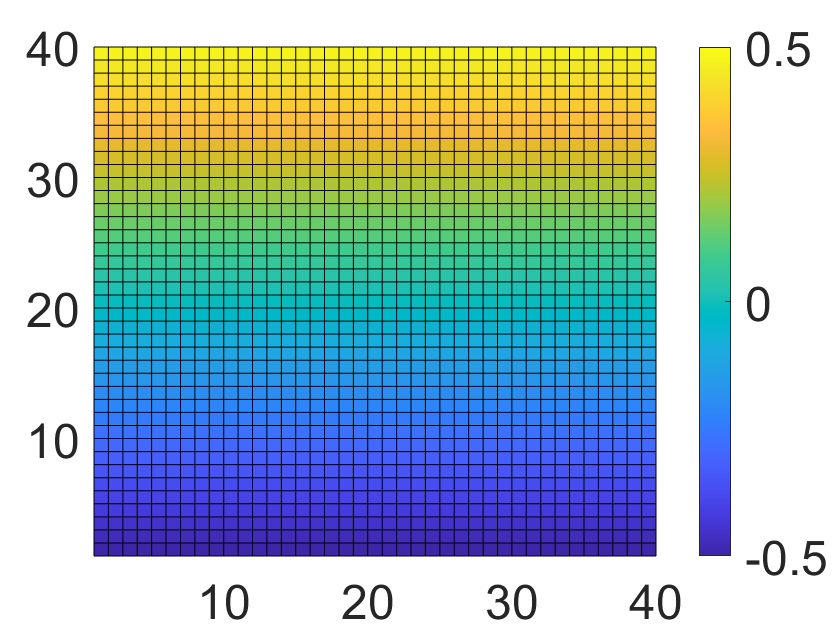}
         \caption{ }
         \label{subfig:GP6 test}
     \end{subfigure}
     \begin{subfigure}{0.329\textwidth}
         \centering
         \includegraphics[width=\textwidth]{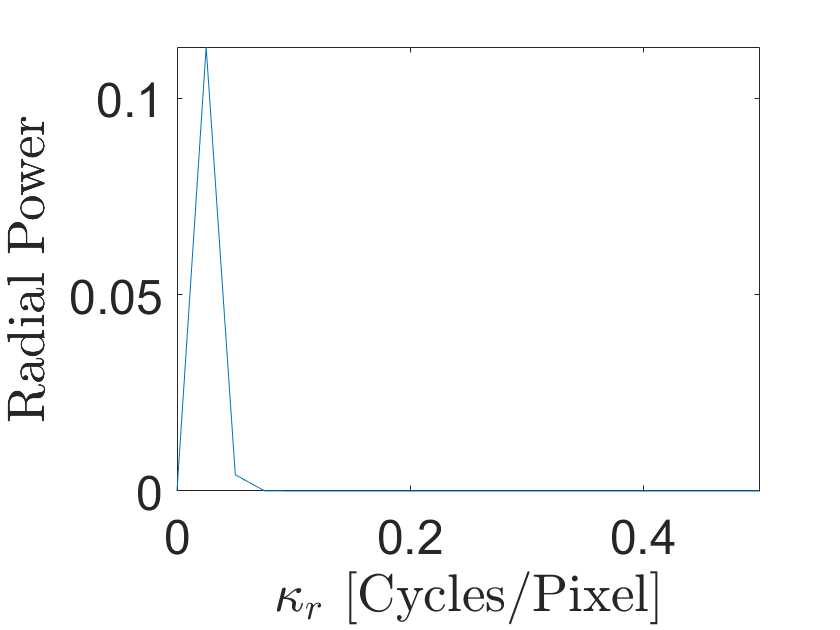}
         \caption{ }
         \label{subfig:freq GP1}
     \end{subfigure}
     \begin{subfigure}{0.329\textwidth}
         \centering
         \includegraphics[width=\textwidth]{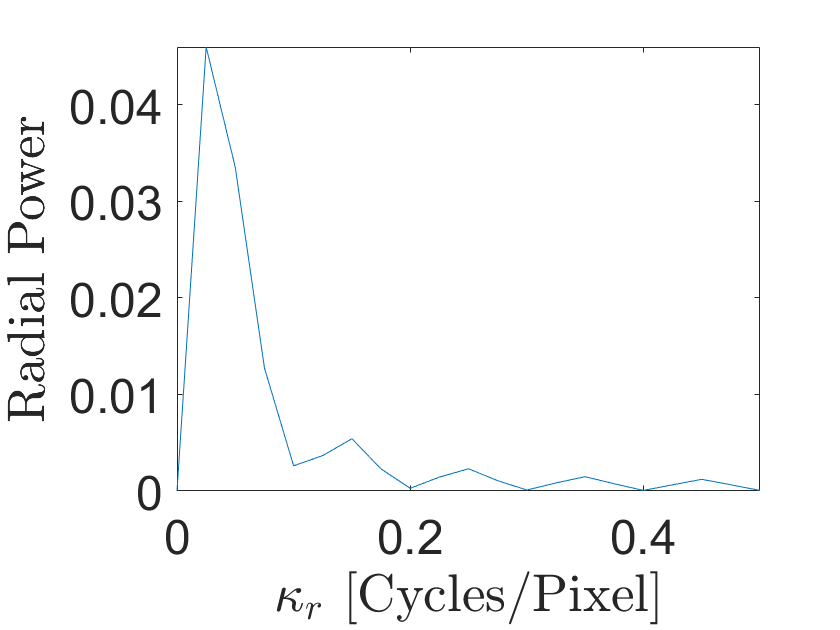}
         \caption{ }
         \label{subfig:freq GP2}
     \end{subfigure}
     \begin{subfigure}{0.329\textwidth}
         \centering
         \includegraphics[width=\textwidth]{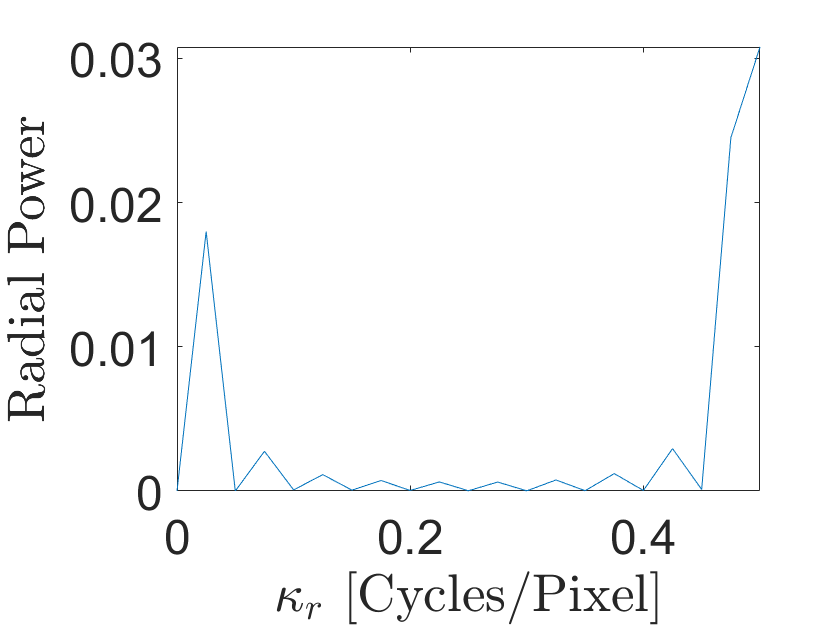}
         \caption{ }
         \label{subfig:freq GP3}
     \end{subfigure}
     \begin{subfigure}{0.329\textwidth}
         \centering
         \includegraphics[width=\textwidth]{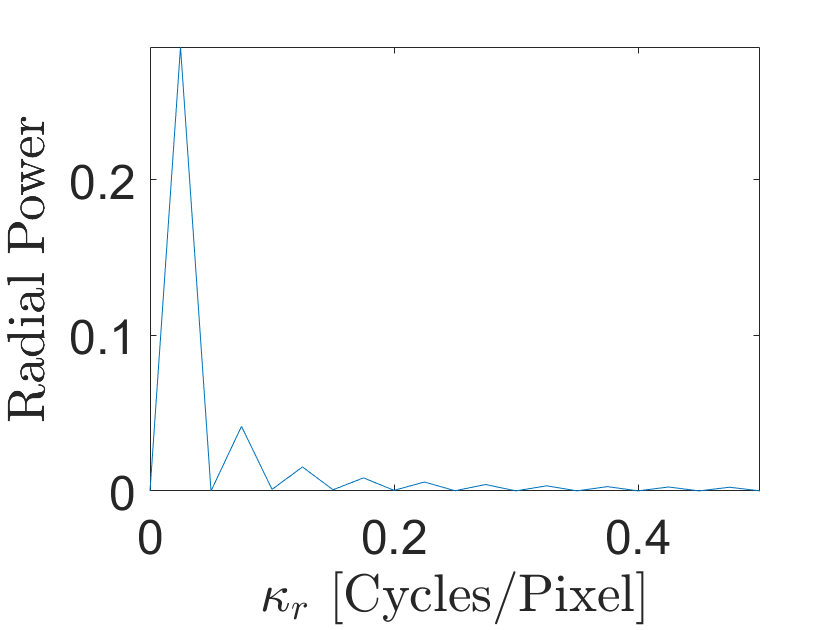}
         \caption{ }
         \label{subfig:freq GP4 }
     \end{subfigure}
     \begin{subfigure}{0.329\textwidth}
         \centering
         \includegraphics[width=\textwidth]{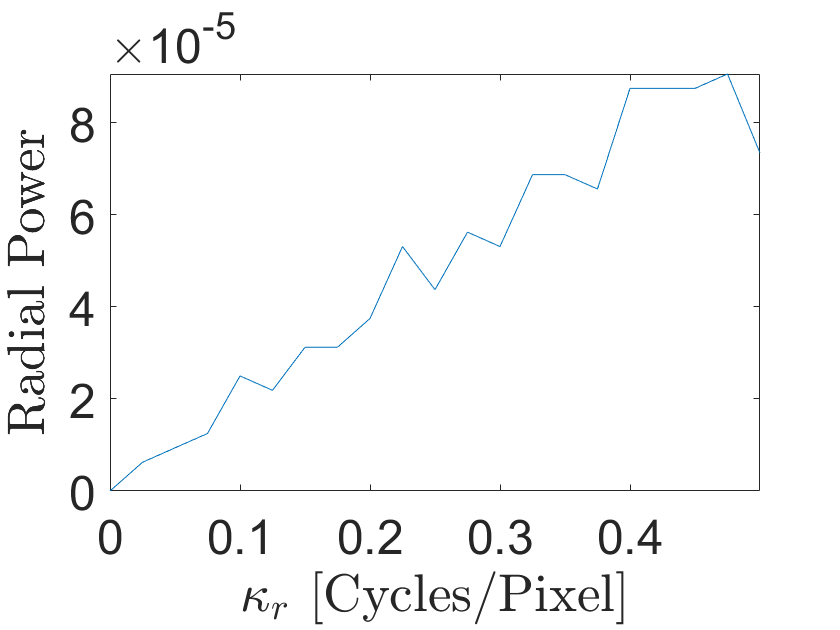}
         \caption{ }
         \label{subfig:freq GP5}
     \end{subfigure}
     \begin{subfigure}{0.329\textwidth}
         \centering
         \includegraphics[width=\textwidth]{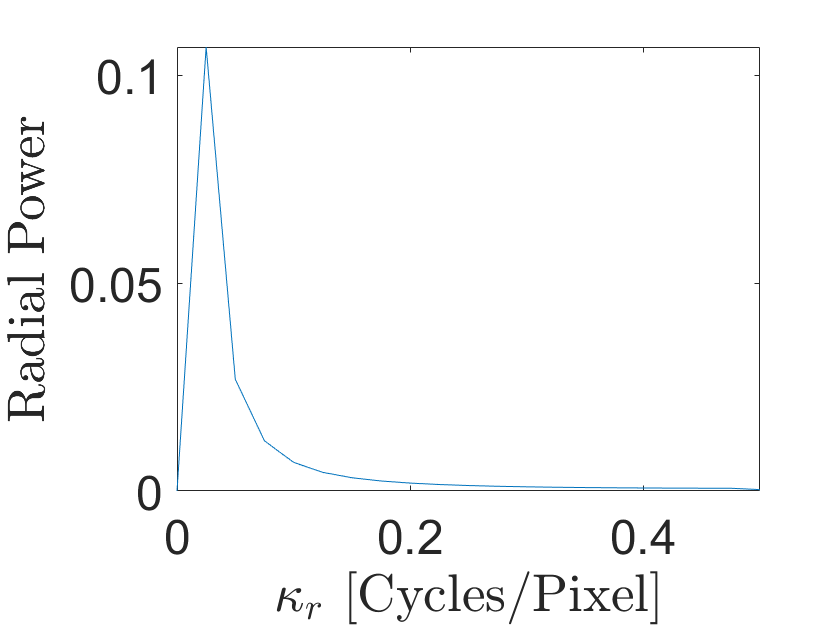}
         \caption{ }
         \label{subfig:freq GP6}
     \end{subfigure}
        \caption{(a)-(f) Desired ghost projection test images. We refer to these as: (a) Gaussian dot, (b) square, (c) dots, (d) large square, (e) small square, and (f) linear gradient. (g)-(l) Corresponding angularly integrated 2-dimensional Fourier power spectra, expressed as a function of dimensionless radial spatial frequency $\kappa_r$.}
        \label{fig:six freq probing graphs}
\end{figure*}

To probe the image norm and power spectrum, we explored the performance of eight test images, beginning with (i) a Gaussian dot, (ii) a square dot and (iii) a series of smaller square dots.  All of the first three cases have the same contrast and image norm.  Relative to the previously mentioned square dot (ii), we further explored (iv) a larger square dot and (v) a smaller square dot.  We also explored (vi) a linear gradient, (vii) the previously mentioned binary resolution chart in Fig.~\ref{subfig: GP1} and (viii) a Gaussian smoothed version of the binary resolution chart performed with a standard deviation of 1.2 pixels. The first six of these images can be seen in Fig.~\ref{fig:six freq probing graphs}a-f. The corresponding power spectra are shown in Fig.~\ref{fig:six freq probing graphs}g-l.

In order to isolate the image parameters from the mask parameters, these images were all ghost projected using the Ni foam master mask with the same set of sub-FOVs being $N=5nm$ in size and sampled in 12-by-8 pixels increments between FOVs in the $x$-by-$y$ axis. Table \ref{Table:ImageParameters} displays (i) the square of the image norm E$[I^2]$, (ii) the dominant power spectrum frequency, (iii) the weighted average, plus-or-minus the standard deviation, of the power spectrum frequencies, (iv) the NNLS ghost projection SNR and pedestal of each of the 8 images, in addition to (v) their performance with the inclusion of Poisson, exposure and translational noise contributions. 

\subsubsection{Image norm}

The image norm $\sqrt{\text{E}[I^2]}$ quantifies the ``amount of signal'' contained in a ghost projection, consistent with our definition of SNR in Eq.~(\ref{eq:SNRdefinition}). The simple notion ``more signal\footnote{By ``signal'', we are referring to $\sqrt{\text{E}[I^2]}$ for the same number of photons being allocated per pixel and are excluding the possibility of increasing the number of photons used to construct the projection as a means of increasing the signal.  However, so long as having an increased number of photons is not an issue for the given application at hand, this should also improve the final SNR.} equals higher final SNR'' is not entirely accurate, as the amount of high frequency power contained in the desired image also impacts a ghost projection's representation characteristics (the weights $w_k$, the delroughness $\mathscr{D}$, etc.) and final SNR. If we compare the test images ``small square'', ``square'' and ``large square'', see Fig.~\ref{fig:six freq probing graphs}, which are 1-by-1, 10-by-10, and 20-by-20 pixel squares respectively, we can observe the final SNR progresses as $0.26 \pm 0.01$, $2.16 \pm 0.1$ and $3.53 \pm 0.18$ which is approximately linear between $\sqrt{\text{E}[I^2]}$ and final SNR. Moreover, we can also compare the ``small square'' image to the ``dots'' image (being 100 spaced small squares). Both have similar power spectrum properties and we can observe an SNR increase from $0.26 \pm 0.01$ to $0.71 \pm 0.03$. 

So, whilst in general ``more signal'' might not linearly lead to ``more final SNR'', it appears approximately true for ghost projections of self-similar images. Whilst not linearly true for images in general, there is still a loose, positive correlation between the amount of signal and final SNR. This suggests that we would want to ``zoom in'' and fill the ghost projection FOV, as much as is reasonably possible, with the features we wish to write. This may be achieved by combining raster scanning with ghost projection, i.e.~creating the ghost projection by isolating the smallest window feasible given the radiant energy and forming a ghost projection within this window before tiling together the entire desired projection.\footnote{This scheme for parallelized ghost projection is analogous to the scheme for parallelized ghost imaging reported by Kingston {\em et al.}~\cite{Kingston2020} in the context of neutron ghost imaging, and Zhang {\em et al.}~\cite{Zhang2022} in the context of x-ray ghost imaging. } On this note, by zooming in, we may also be able to minimize the resolution necessary to resolve the masks and desired image which may impact the number of sub-FOVs the ghost projection picks out since this is typically of the order of the resolution, $N' \approx nm$. A smaller number of masks $N'$ typically leads to a smaller pedestal which should further improve the ghost projection's robustness to noise inclusions. For example, if it were reasonable to do so, a ghost projection on a $20 \times 20$ grid will outperform a ghost projection on a $40 \times 40$ grid -- the difficulty being if it is indeed ``reasonable to do so''. If there are unresolved features in the mask at the coarser resolution, then these will contaminate the final ghost projection on a sub-pixel scale. 

\subsubsection{Image power spectrum}

The image power spectrum gives us an indication of the distribution of the signal for a given resolution and FOV. Consider the Gaussian dot, square dot and dots images which all have the same image norm, $\sqrt{\text{E}[I^2]} = 0.242$, but have different mode, mean and standard deviations in terms of power spectrum frequencies (image distribution), as seen in Table \ref{Table:ImageParameters}. We observe the trend that images characterized by low-frequency power tend to, on average, achieve a higher final SNR. Seemingly, if the desired image contains a significant proportion of high-frequency power relative to what is present in the masks, the reconstruction has to ``work harder'' to represent the desired image, which can be observed in an increase in the ghost projection pedestal -- cf.~a pedestal of 4.8, 45.6 and 331.4 for the Gaussian dot, square dot and dots, respectively.

Similarly, we might compare the binary resolution chart and the Gaussian smoothed binary resolution chart. By smoothing off the sharp edges, the NNLS ghost projection pedestal shrinks from 121.7 to 13.7 and the final SNR with Poisson, exposure and translational noise increases from $2.67 \pm 0.11$ to $4.10 \pm 0.28$. In contemplating this final value, we should keep in mind that the Gaussian smoothing could also be considered a source of noise, whereby we have assumed that the smoothed version is the desired outcome. Moreover, we might also note that due to the changed reconstruction of the smoothed version, the SNR with Poisson noise only and SNR with translational noise only markedly increase from the unsmoothed version -- that is, a factor of 2.3 and 6.7, respectively. We might understand this improvement as coming from a decrease in the weights used and decrease in the delroughness of the set of masks selected. In fact, the limiting noise contribution preventing the smoothed binary resolution chart from reaching an even higher final SNR is the exposure noise, where this actually decreases from the unsmoothed to the smoothed case, $5.62 \pm 0.48$ to $4.60 \pm 0.41$. With reference to Eq.~(\ref{eq: analytical exposure noise prediction}), since $\sigma_w$ is unchanged and $N'$ decreases from 1581 to 1566, we must assume that the limiting factor is the variance of the mask transmission values Var$[R]$. This suggests that, should we want to improve the final SNR of the Gaussian-smoothed binary resolution chart, focusing on reducing Var$[R]$ would be the best course of action. 

One final point of interest with respect to image power spectra is that for all of the images that do not contain sharp edges---such as the Gaussian dot, linear gradient and Gaussian-smoothed binary resolution chart---we saw remarkable robustness to translational noise. These images obtained an SNR, with translational noise only, of $60.16 \pm 2.88$, $43.17 \pm 2.12$ and $30.28 \pm 1.70$, respectively. Although it is often counter-productive to set out to produce a smoothed projection, this preliminary investigation suggests that it could prove fruitful for improving the overall SNR to dial back the sharpness to a value which is just above the tolerable sharpness given a particular application. 


\begin{table*}[ht!]
\renewcommand{\arraystretch}{1.2}
\begin{tabular}{|c|c|c|c|c|c|c|c|c|}
\hline
Image                                                                                                    & Gaussian                                                       & Square                                                       & Dots                                                         & \begin{tabular}[c]{@{}c@{}}Large\\ Square\end{tabular}      & \begin{tabular}[c]{@{}c@{}}Small\\ Square\end{tabular}       & \begin{tabular}[c]{@{}c@{}}Linear\\ Gradient\end{tabular}    & \begin{tabular}[c]{@{}c@{}}Binary \\ Res. Chart\end{tabular} & \begin{tabular}[c]{@{}c@{}}Gaus. Smoothed\\ Binary \\ Res. Chart\end{tabular} \\ \hline
E$[I^2]$                                                                                                 & 0.0586                                                      & 0.0586                                                       & 0.0586                                                       & 0.1875                                                      & 0.0006                                                       & 0.0876                                                       & 0.2097                                                       & 0.1209                                                                        \\ \hline
\begin{tabular}[c]{@{}c@{}}Peak Freq.\\ {[}cyc./pix.{]}\end{tabular}                                     & 0.025                                                       & 0.025                                                        & 0.500                                                        & 0.025                                                       & 0.475                                                        & 0.025                                                        & 0.075                                                        & 0.075                                                                         \\ \hline
\begin{tabular}[c]{@{}c@{}}E$[\kappa_r] \pm \sqrt{\text{Var}[\kappa_r]}$ \\ {[}cyc./pix.{]}\end{tabular} & \begin{tabular}[c]{@{}c@{}}0.026\\ $\pm 0.005$\end{tabular} & \begin{tabular}[c]{@{}c@{}}0.079 \\ $\pm 0.090$\end{tabular} & \begin{tabular}[c]{@{}c@{}}0.359 \\ $\pm 0.198$\end{tabular} & \begin{tabular}[c]{@{}c@{}}0.056\\ $\pm 0.077$\end{tabular} & \begin{tabular}[c]{@{}c@{}}0.337 \\ $\pm 0.120$\end{tabular} & \begin{tabular}[c]{@{}c@{}}0.063 \\ $\pm 0.080$\end{tabular} & \begin{tabular}[c]{@{}c@{}}0.119\\ $\pm 0.105$\end{tabular}  & \begin{tabular}[c]{@{}c@{}}0.067 \\ $\pm 0.036$\end{tabular}                  \\ \hline
NNLS SNR                                                                                                      & $1.40 \times 10^8$                                          & $1.65 \times 10^8$                                           & $1.83 \times 10^8$                                           & $2.67 \times 10^8$                                          & $1.38 \times 10^7$                                           & $2.13 \times 10^8$                                           & $3.34 \times 10^8$                                           & $2.73 \times 10^8$                                                            \\ \hline
$\bar{P}$                                                                                                & 4.80                                                        & 45.6                                                         & 331.4                                                        & 60.1                                                        & 24.8                                                         & 8.15                                                         & $121.7$                                                      & 13.7                                                                          \\ \hline
$N'$                                                                                                     & 1563                                                        & 1575                                                         & 1581                                                         & 1576                                                        & 1574                                                         & 1564                                                         & $1581$                                                       & 1566                                                                          \\ \hline
\begin{tabular}[c]{@{}c@{}}SNR \\ w/ Poisson\end{tabular}                                                & $11.08 \pm 0.20$                                            & $3.59 \pm 0.07$                                              & $1.33 \pm 0.03$                                              & $5.57 \pm 0.09$                                             & $0.50 \pm 0.01$                                              & $10.42 \pm 0.16$                                             & $4.15 \pm 0.07$                                              & $9.40 \pm 0.16$                                                               \\ \hline
\begin{tabular}[c]{@{}c@{}}SNR \\ w/ Exposure\end{tabular}                                               & $3.61 \pm 0.29$                                             & $3.03 \pm 0.29$                                              & $2.90 \pm 0.25$                                              & $5.30 \pm 0.41$                                             & $0.32 \pm 0.02$                                              & $4.13 \pm 0.35$                                              & $5.62 \pm 0.48$                                              & $4.60 \pm 0.41$                                                               \\ \hline
\begin{tabular}[c]{@{}c@{}}SNR \\ w/ Translational\end{tabular}                                          & $60.16 \pm 2.88$                                            & $6.35 \pm 0.33$                                              & $0.87 \pm 0.05$                                              & $8.69 \pm 0.49$                                             & $1.19 \pm 0.06$                                              & $43.17 \pm 2.12$                                             & $4.52 \pm 0.21$                                              & $30.28 \pm 1.70$                                                              \\ \hline
SNR w/ All                                                                                               & $3.31 \pm 0.22$                                             & $2.16 \pm 0.10$                                              & $0.71 \pm 0.03$                                              & $3.53 \pm 0.18$                                             & $0.26 \pm 0.01$                                              & $3.75 \pm 0.22$                                              & $2.67 \pm 0.11$                                              & $4.10 \pm 0.28$                                                               \\ \hline
\end{tabular}
\caption{Investigation into the effect of image parameters on the ghost projection NNLS SNR, pedestal $\bar{P}$ and SNR with all contributions of noise. This was performed with $N = 5nm$ Ni foam masks, sampled at 12-by-8 intervals along the $x$ and $y$-axis, respectively. Poisson noise, exposure noise and translational noise were for the conditions $\lambda = 10\,000$, $\sigma_w = 1/100$ and $\sigma_{ij} = 1/10$, respectively. The Gaussian smoothing applied to the binary resolution chart was with a standard deviation of 1.2 pixels.}
\label{Table:ImageParameters}
\end{table*}

\subsection{Source parameters}

Here we explore the ability of numerically optimized ghost projection to adapt to non-uniform illumination. To model this, we apply a Gaussian profile over the $40 \times 40$ mask FOVs to represent the lower flux generated further from the beam center and seek a representation that compensates for these relatively dimmed boundaries. The Gaussian profile employed, expressed as a transmission coefficient, is
\begin{align}
    s_{ij} = \exp \left( - \alpha \left[ \left(x_{i}-x_c\right)^2 + \left(y_{j}-y_c\right)^2 \right] \right),
\end{align}
where $\alpha$ is the profile decay parameter that controls the transmission experienced in the boundaries of the FOV, $x_i$ and $y_j$ index the spatial domain, and $(x_c,y_c)$ are the coordinates of the center of the FOV. For brevity, we quote the effect of the Gaussian profile as the transmission experienced in the corner of the FOV, e.g.~a 95\% Gaussian profile correlates to 95\% of the uniform beam being transmitted in the corner of the FOV. Here we explored increasing non-uniformity, with Gaussian profiles 90\%, 80\% and 70\% for the Ni foam mask, with $N=5nm$ sampled in strides of 12-by-8 pixels, used to reconstruct the binary resolution chart. The results of this investigation are shown in Table \ref{table:Source Profile}. In this, we can observe that the reconstruction appears to have to ``work harder'' to compensate for the relatively dimmed boundaries in the 90\% and 80\% transmission cases, as can be seen in the increased pedestal. This cannot continue ad infinitum however, and in the 70\% transmission case, we see that the NNLS reconstruction is only able to reach an initial SNR of 3.65. Admittedly, this is quite a robust 3.65 which outperforms the other cases considered here, including that of no source attenuation, in terms of final SNR. This result should not be overly extrapolated, though.  If the number of photons used in the ghost projection were to be increased, or the exposure noise and translational noise parameters $\sigma_w$ and $\sigma_{ij}$ were to be reduced, then the 70\% transmission case can only ever hope to achieve a final SNR in the neighborhood of its initial SNR of 3.65, whereas the others are not limited in the same fashion. Moreover, this is yet another example of how NNLS optimization of the noise-free ghost projection does not necessarily yield a reconstruction that is most robust to noise inclusions.  Future effort focused on the optimization algorithm might therefore be a source of significant additional improvements in the final SNR. 

\begin{table*}[ht!]
\renewcommand{\arraystretch}{1.2}
\begin{tabular}{|c|c|c|c|c|}
\hline
Gaussian Profile     & 100\%              & 90\%               & 80\%               & 70\%            \\ \hline
NNLS SNR             & $3.34 \times 10^8$ & $2.80 \times 10^8$ & $2.48 \times 10^9$ & 3.65            \\ \hline
$\bar{P}$            & $121.7$            & 132.1              & 413.8              & 33.79           \\ \hline
$N'$                 & $1581$             & 1580               & 1596               & 660             \\ \hline
SNR w/ Poisson       & $4.15 \pm 0.07$    & $3.97 \pm 0.07$    & $2.25 \pm 0.04$    & $3.31 \pm 0.03$ \\ \hline
SNR w/ Exposure      & $5.62 \pm 0.48$    & $5.66 \pm 0.44$    & $5.81 \pm 0.48$    & $3.41 \pm 0.04$ \\ \hline
SNR w/ Translational & $4.52 \pm 0.21$    & $4.10 \pm 0.21$    & $1.36 \pm 0.07$    & $3.42 \pm 0.02$ \\ \hline
SNR w/ All           & $2.67 \pm 0.11$    & $2.56 \pm 0.10$    & $1.14 \pm 0.05$    & $2.98 \pm 0.05$ \\ \hline
\end{tabular}
\caption{Ghost projections with Gaussian transmission profiles applied to the mask FOVs where the value quoted is the transmission percentage in the corners of the FOV. The ghost projections were performed with the Ni foam master mask, where $N = 5nm$, strided 12-by-8 in the $x$ and $y$-axis respectively. Poisson, exposure, translational noise were for the conditions $\lambda = 10\,000$, $\sigma_w = 1/100$ and $\sigma_{ij} = 1/10$, respectively.}
\label{table:Source Profile}
\end{table*}

\subsection{Prescribed pedestal}

Rather than moving the pedestal to the left-hand side in our derivation of the NNLS approach (see Eq.~(\ref{eq:Move Pedestal to LHS})), we can leave it on the right-hand side and simply enforce an arbitrarily selected value. The motivation for doing this is that our previous construction often found an excellent starting SNR, but had no incentive to minimize the pedestal, weights or number of filtered masks, which are significant factors in determining how robust the ghost projection is to Poisson, exposure and translational noise. Whilst arbitrarily selecting a pedestal is not the optimal approach, we can gain a preliminary indication for what kinds of SNR improvements are possible from our previously obtained baseline values. 

In Table \ref{table:EnforcedPedestals}, we can observe the result of enforcing a pedestal that is 1, 2, 5, 10, 20, 50 and 100 times the contrast of the desired image. For all of the prescribed pedestals, the NNLS ghost projection was performed with $5nm$ Ni foam masks that were sampled systematically in strides of 12-by-8 pixels. For the prescribed pedestal of 1 and 5, these cases were also ghost projected with $50nm$ Ni foam sub-FOVs that were sampled systematically in steps of 1-by-1 pixels. The desired image employed here was the binary resolution chart, Fig.~\ref{subfig: GP1}. Given that the binary resolution chart has sharp edges and fine-scale features, the ghost projection generally has to work quite hard to create these characteristics. In our baseline investigation of ghost projecting the binary resolution chart with $5nm$ Ni foam masks, the pedestal was 121.7. With this in mind, we can foresee that enforcing a pedestal of 1, 2, 5, 10 and 20 are all quite severe restrictions. This is reflected in the corresponding NNLS SNRs obtained: 1.34, 1.67, 2.48, 3.32 and 4.89. Neither did it seem to help much by increasing the number of masks available to the NNLS optimizer by a factor of 10, as was done for the enforced pedestal of 1 and 5. These cases had NNLS SNRs of 1.37 and 2.80, respectively. It was not until an enforced pedestal of 50 times the projection contrast that we saw the benefits of a reduced pedestal profitably being traded in for some of the baseline NNLS SNR. In this case, we saw an initial NNLS SNR of 15.85 which was eroded down to $3.93 \pm 0.08$ once all noise inclusions were accounted for (which we might compare to $2.67 \pm 0.11$ for the non-prescribed pedestal case). By the time we prescribe a pedestal of 100, we see that we have overshot the optimal trade-off, where the initial NNLS SNR is on the order of $10^8$ and the final SNR is lower than the previously enforced pedestal.

\begin{table*}[ht!]
\centering
\renewcommand{\arraystretch}{1.2}
\begin{tabular}{|c|cc|c|cc|c|c|c|c|}
\hline
Enforced $\bar{P}$                                              & \multicolumn{2}{c|}{1}                                 & 2               & \multicolumn{2}{c|}{5}                                 & 10              & 20              & 50              & 100                \\ \hline
$N$                                                             & \multicolumn{1}{c|}{$5nm$}          & $50nm$          & $5nm$           & \multicolumn{1}{c|}{$5nm$}           & $50nm$          & $5nm$           & $5nm$           & $5nm$           & $5nm$              \\ \hline
Striding                                                        & \multicolumn{1}{c|}{12-by-8}         & 1-by-1          & 12-by-8         & \multicolumn{1}{c|}{12-by-8}         & 1-by-1          & 12-by-8         & 12-by-8         & 12-by-8         & 12-by-8            \\ \hline
NNLS SNR                                                        & \multicolumn{1}{c|}{1.34}            & 1.37            & 1.67            & \multicolumn{1}{c|}{2.48}            & 2.80            & 3.32            & 4.89            & 15.85           & $1.48 \times 10^8$ \\ \hline
$N'$                                                            & \multicolumn{1}{c|}{30}              & 41              & 65              & \multicolumn{1}{c|}{212}             & 291             & 440             & 748             & 1319            & 1582               \\ \hline
\begin{tabular}[c]{@{}c@{}}SNR w/ \\ Poisson\end{tabular}       & \multicolumn{1}{c|}{$1.34 \pm 0.00$} & $1.37 \pm 0.00$ & $1.67 \pm 0.00$ & \multicolumn{1}{c|}{$2.46 \pm 0.01$} & $2.78 \pm 0.01$ & $3.24 \pm 0.02$ & $4.42 \pm 0.04$ & $6.00 \pm 0.11$ & $4.58 \pm 0.08$    \\ \hline
\begin{tabular}[c]{@{}c@{}}SNR w/ \\ Exposure\end{tabular}      & \multicolumn{1}{c|}{$1.34 \pm 0.01$} & $1.38 \pm 0.01$ & $1.67 \pm 0.01$ & \multicolumn{1}{c|}{$2.46 \pm 0.01$} & $2.76 \pm 0.02$ & $3.18 \pm 0.03$ & $4.21 \pm 0.10$ & $5.56 \pm 0.41$ & $5.48 \pm 0.48$    \\ \hline
\begin{tabular}[c]{@{}c@{}}SNR w/ \\ Translational\end{tabular} & \multicolumn{1}{c|}{$1.30 \pm 0.00$} & $1.33 \pm 0.00$ & $1.60 \pm 0.00$ & \multicolumn{1}{c|}{$2.36 \pm 0.00$} & $2.65 \pm 0.00$ & $3.13 \pm 0.01$ & $4.46 \pm 0.02$ & $8.37 \pm 0.31$ & $5.59 \pm 0.29$    \\ \hline
\begin{tabular}[c]{@{}c@{}}SNR w/ \\ All\end{tabular}           & \multicolumn{1}{c|}{$1.30 \pm 0.01$} & $1.33 \pm 0.01$ & $1.60 \pm 0.01$ & \multicolumn{1}{c|}{$2.33 \pm 0.02$} & $2.60 \pm 0.02$ & $2.95 \pm 0.03$ & $3.66 \pm 0.08$ & $3.93 \pm 0.15$ & $2.97 \pm 0.13$    \\ \hline
\end{tabular}
\caption{SNR and number of non-zero weights for enforced pedestal for NNLS ghost projection with the Ni foam mask. The Poisson, exposure and translational noise are for the conditions $\lambda = 10\,000$, $\sigma_w = 1/100$ and $\sigma_{ij} = 1/10$, respectively. }
\label{table:EnforcedPedestals}
\end{table*}

\subsection{Estimate of experiment duration via traveling salesperson problem} \label{Old AppB}

For a ghost-projection scenario that involves a single mask, suppose that a suitable numerical optimization process has specified a given set of $N'$ transverse mask translations 
\begin{equation}
\mathcal{S}\equiv\{(\delta x_j, \delta y_j)\}, \quad \text{for } j=1,2,\cdots,N' 
\end{equation}
that are to be employed.  This set of transverse mask positions can be illuminated in any temporal sequence.  To minimize the total experiment duration, it will typically be reasonable to choose a particular temporal sequence of mask positions which minimizes the total distance that is traveled, by the transverse-translation stage upon which the mask is fixed.  This minimization problem is equivalent to the famous ``traveling salesperson problem'' \cite{Armour1983} of finding the shortest path which connects all of the points in $\mathcal{S}$.  

To estimate the duration of a ghost projection experiment, we separately consider the scan time $t_s$ and exposure time $t_e$. We estimate the scan time via
\begin{align}
    t_{s} = \frac{L}{\bar{v}},
\end{align}
where $L$ is the length of the scan path and $\bar{v}$ is the average speed of the mask-translation stage. We estimate the exposure time via
\begin{align}
    t_e = \frac{\lambda J^k w_k}{ \Phi }, 
\end{align}
where $\lambda$ is our desired number of photons per pixel used to create the projection contrast and $\Phi$ is the source flux in units [photons/pixel/time]. Note, the reasonableness of our exposure noise parameter $\sigma_w$ is directly related to source intensity. In some circumstances it may be sensible to dim the source and thus reduce the requirement for a rapidly responding shutter.

\begin{figure*}[ht!]
\centering
     \begin{subfigure}{0.3\textwidth}
         \centering
         \includegraphics[width=\textwidth]{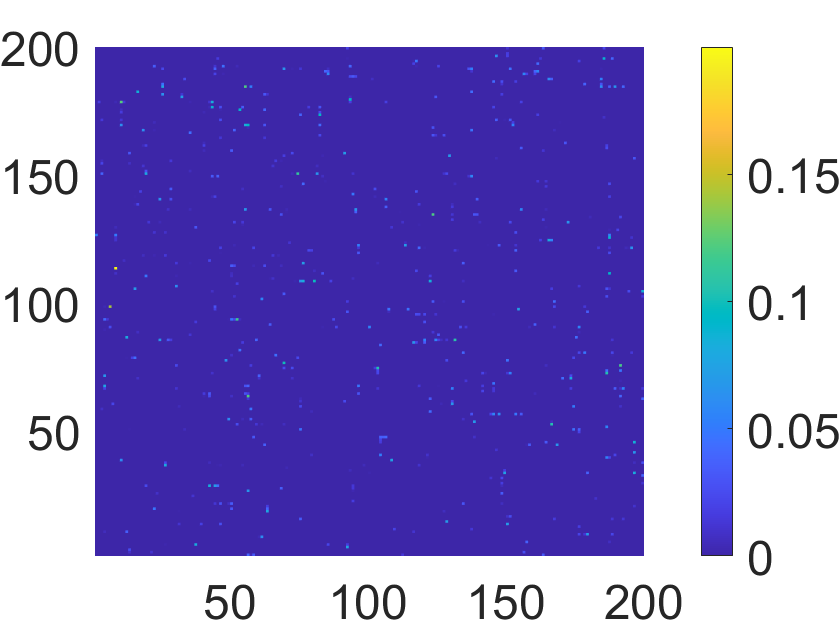}
         \caption{ }
         \label{subfig: WD1}
     \end{subfigure}
     \begin{subfigure}{0.3\textwidth}
         \centering
         \includegraphics[width=\textwidth]{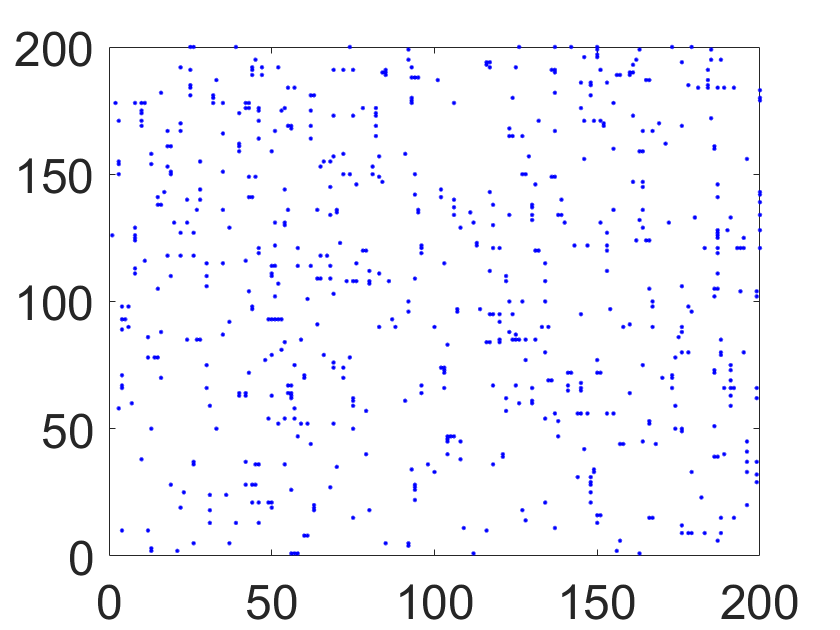}
         \caption{ }
         \label{subfig: WD2}
     \end{subfigure}
     \begin{subfigure}{0.3\textwidth}
         \centering
         \includegraphics[width=\textwidth]{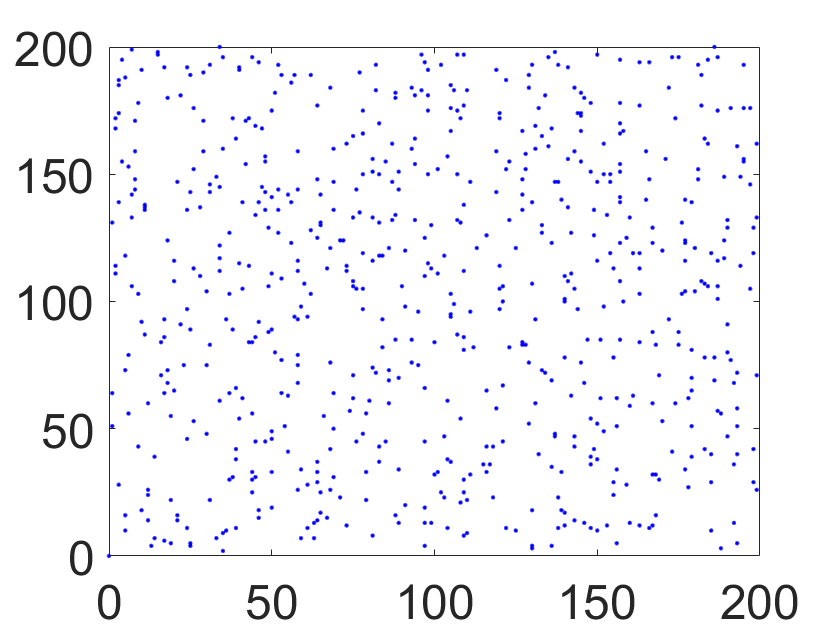}
         \caption{ }
         \label{subfig: WD3}
     \end{subfigure}
\caption{(a) Plot of the NNLS weighting coefficients as a function of transverse offset for the Ni foam mask used to ghost project the Gaussian smoothed ($\sigma = 0.5$ pixels) binary resolution chart with an enforced pedestal of 10 and $N = 25nm$ sub-FOVs stepped in 1-by-1 pixel increments. This had an initial NNLS SNR of $6.80$ and the number of sub-FOVs selected was 672. Once noise was included, the SNR dropped to $4.70 \pm 0.14$ for the noise parameters $\lambda = 10\,000$, $\sigma_w = 1/100$ and $\sigma_{ij} = 1/10$. (b) A binary version of the NNLS weights displayed in (a). (c) 672 points chosen uniformly at random on a $200 \times 200$ integer grid. In comparison to (b) we can observe some similarities in the uniform sampling of mask displacements, although, within (b) there appears to be locations of horizontal and vertical sampling of the Ni foam mask which we speculate is the utilization of constructive and destructive interference of the mask speckles to make vertical and horizontal features, respectively.}
\label{Fig:WeightsDistribution}
\end{figure*}

For the purposes of estimating the duration of an experiment, we consider ghost projecting the Gaussian smoothed binary resolution chart ($\sigma = 0.5$ pixels) with an enforced pedestal of 10 made with $N = 25 nm$ Ni foam sub-FOVs, sampled in 1-by-1 pixel increments. The NNLS weights, expressed as a function of transverse offset, can be observed in Fig.~\ref{subfig: WD1} as well as a binary version in Fig.~\ref{subfig: WD2}. This ghost projection had an initial NNLS SNR of $6.80$ and the number of sub-FOVs selected was 672. Once noise was included, the SNR dropped to $4.70 \pm 0.14$ for the noise parameters $\lambda = 10\,000$, $\sigma_w = 1/100$ and $\sigma_{ij} = 1/10$.

As an interesting and topical note, compare the NNLS weights distribution in Fig.~\ref{subfig: WD2}, to the distribution of 672 points sampled uniformly at random from the same $200 \times 200$ integer grid, in Fig.~\ref{subfig: WD3}. By eye, there are some similarities between the two in terms of uniform sampling but it would be a mis-characterization of the distribution of NNLS weights to say they are the same. In Fig.~\ref{subfig: WD2}, we can also observe some evidence of the NNLS optimizer utilizing constructive and destructive interference by the presence of relatively small horizontal and vertical features in the $w_k$ locations. That is, in order to create a relatively sharp horizontal edge in projection, we might scan a section of the mask vertically to create this, or vice versa for a relatively sharp vertical edge.

\begin{figure*}[ht!]
    \centering
    \begin{subfigure}{0.35\textwidth}
         \centering
         \includegraphics[width=0.92\textwidth, clip,trim={4.5cm 1cm 26.2cm 0cm}]{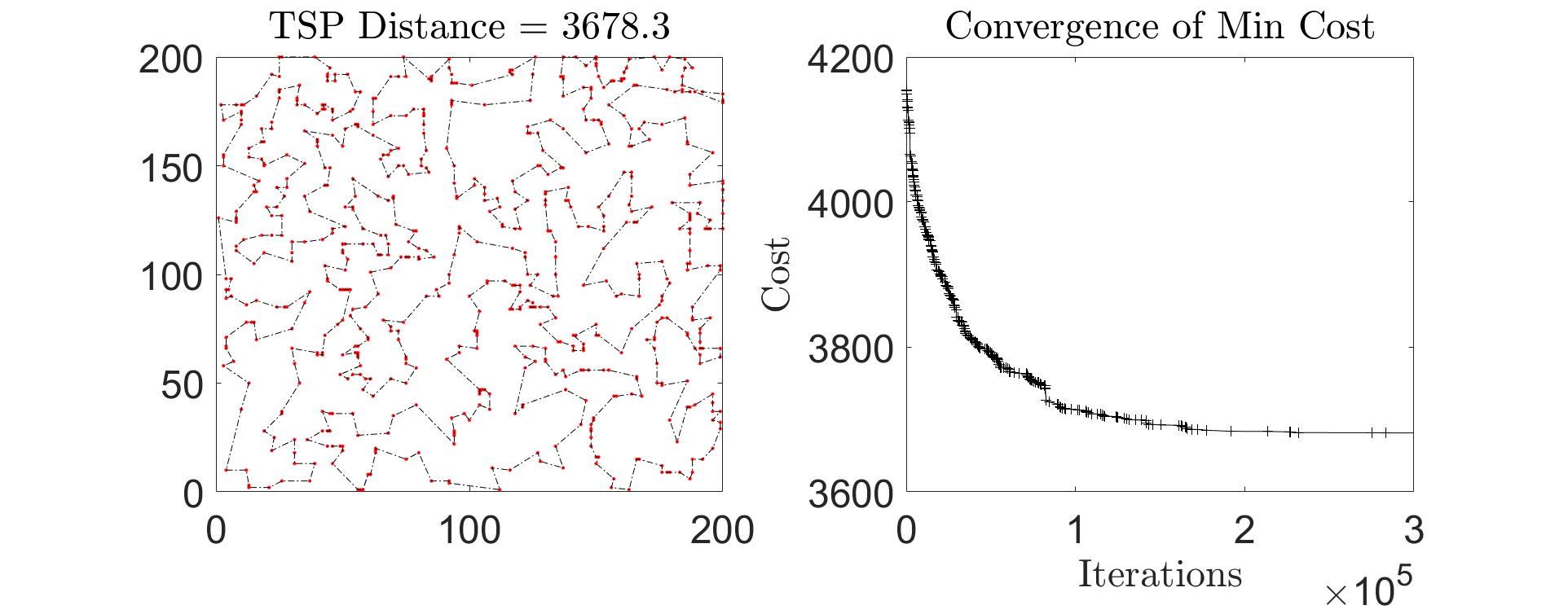}
         \caption{ }
         \label{subfig:TSP path}
     \end{subfigure}
     \begin{subfigure}{0.35\textwidth}
         \centering
         \includegraphics[width=1.1\textwidth,clip,trim={24.6cm 0cm 1cm 0cm}]{TSP_solution_N2-2.png}
         \caption{ }
         \label{subfig:TSP convergence}
     \end{subfigure}
        \caption{ (a) Heuristically optimized traveling salesperson problem for routing of the Ni foam scan path to ghost project the Gaussian smoothed ($\sigma = 0.5$ pixels) binary resolution chart with an enforced pedestal of 10 and $N = 25nm$ sub-FOVs stepped in 1-by-1 pixel increments. The path distance is in units of pixels. (b) Convergence of the heuristic optimizer as a function of iteration number.   }
        \label{fig:TSP}
\end{figure*}

With our binarized, transverse distribution of NNLS weights, we can now treat this as a traveling salesperson problem (TSP) \cite{Armour1983} to determine the length of the scan path. As mentioned earlier, the TSP seeks the shortest possible path joining a specified set of points in the plane.  In its usual form, the TSP path is closed, i.e.~the salesperson starts and ends at the same location.  For application of the TSP in the context of ghost projection, the path may be open rather than closed.  However, when the number of mask positions is very large compared to unity, there will typically be a negligible difference between open and closed TSP paths, for the purposes of approximating the total scan time for a ghost-projection experiment. Using the \texttt{MATLAB} function $\mathtt{tsp\_ga\_basic}$ developed by Robert Rich \cite{RobertRichMATLAB}, an estimate for the optimum scan path was obtained as 3\,678.3 pixels,  see Fig.~\ref{fig:TSP}. Given the Ni foam is 30$\mu$m/pixel, this gives the scan path to be 110.35mm. Consulting values for precision x-y-stage speeds, Kohzu Precision Co.~advertises x-y stages that have maximum speeds of 5-50mm/s for stages that have a motion in the range of $\pm$30mm. Taking an average traversal speed of 1mm/s, we estimate the scan time to be 110.35 seconds.

Moving onto the exposure time, suppose we want to allocate $\lambda = 10\,000$ photons into the contrast and 10 times that into the pedestal. To achieve this, we would want to allocate a total of $\lambda J^k w_k = 189\,300$ photons per pixel. Keeping in mind that an exposure time is $t_k = \lambda w_k / \Phi \pm \lambda \sigma_w / \Phi$, it might be best to enforce a source intensity based on shutter speed tolerance $\sigma_w$. For the relatively high energy regime of hard x-rays and beyond, perhaps the shutter can only reliably open and close to within $1/100$ of one second. With this value, we have $\lambda \sigma_w / \Phi = 1/100$, or $\Phi = 10\,000$ photons/pixel/second. This yields an exposure time of 18.93 seconds. For the sake of demonstration, supposing our shutter had a tolerance of $1/2$ a second, we might enforce a source flux of $\Phi = 200$ photons/pixel/second and expect an exposure time of 946.5s (or nearly 16 minutes).

Altogether then, the ghost projection experiment duration, being the scan time $t_s$ plus the exposure time $t_e$, would be on the order of $110 + 20 = 130$ seconds. This assumes that we want a contrast of $10\,000$ photons and can achieve a shutter speed tolerance of $1/100$s. Further, assuming a repeatability in transverse positioning of $1/10$ pixel, we can expect an SNR of $4.70 \pm 0.14$ and a pedestal of $100\,000$ photons. Supposing we wanted to speed up this process, the dominant time consumer was traversal time. We could, for example, eliminate this time by way of multiple consecutive masks and constraining the numerical optimizer to selecting a continuous exposure trajectory. We may also seek to increase the source intensity to reduce the exposure time contribution. Alternatively, we could go in the opposite direction and trade in time taken to complete the ghost projection experiment in order to achieve a higher SNR. This could be achieved by (i) dimming the source such that, for the same shutter speed, the exposure noise is reduced, (ii) using a reconstruction that has a higher initial SNR (but this will likely also have more filtered members $N'$ and thus a longer scan path and higher pedestal associated with it), and/or (iii) we could simply allocate more photons to create the contrast (but at the cost of also adding to the pedestal). It is also worth noting that these time estimates exclude the time taken to measure the masks, numerically optimize the reconstruction, and numerically optimize the routing routine. The first of these is a once-off investment per ghost projection master mask. The latter two are once-off investments per desired projection-image.

\subsection{General guidelines}
We conclude this section with some rules-of-thumb, or guidelines, that have become evident during the above investigations on practical schemes for ghost projection by simulation. We also subsequently adopt our own guidelines in one final simulation example, to demonstrate the kinds of improvements possible. Our guidelines for practical ghost projection are as follows:
\begin{enumerate}
    \item The mask should preferably have low overall transmission, i.e.~low $\text{E}[R]$, as this reduces one contributing factor to the size of the pedestal $\bar{P} = \text{E}[w]\text{E}[R]N'$ and the corresponding contribution to Poisson noise from the pedestal (see Eq.~(\ref{eq:Poisson noise})).
    \item If Poisson noise is the dominant contribution to the overall ghost projection noise, as might be observable in the ghost projection as spatially localized noise that is sub-speckle sized, consider (i)  enforcing a small pedestal in the NNLS optimizer or (ii) using a specialized optimizer that balances a reduced pedestal against random-mask representation error. If the application allows it, increasing the number of photons should also reduce Poisson noise contributions.
    \item If exposure noise is the dominant contribution to the noise, as may be observable in the ghost projection as noise similar in appearance to the speckles, then the speckle contrast in the ghost-projection masks should be minimized, i.e.~$\text{Var}[R]$ should be reduced, as the exposure noise grows according to $\sqrt{\sigma_{w}^2N'\text{Var}[R]}$. One could also reduce exposure noise by a further two methods: (i) investing in an improved shutter that reduces the variance in an exposure window, $\sigma_w^2$, or (ii) dimming the source such that for the same exposure variance, it is relatively reduced. 
    \item If translational noise is the dominant contribution to the noise, as may also be observable in the ghost projection as noise similar in appearance to the speckles, then reducing the reconstruction exposure times $w_k$ and the mask delroughness, Eq.~(\ref{eq:Delroughness definition}), should be the focus, as translational noise is proportional to $\sigma_{ij}   w_k \mathscr{D}^k$. Furthermore, one could also reduce translational noise by investing in an improved-precision translational stage that has highly repeatable positioning, thus reducing $\sigma_{ij}$.
    \item The resolution of the ghost projection is limited by the highest spatial frequencies, which are present in the mask power spectra to a non-negligible degree. So, should one want a sharp ghost projection, the mask speckles should have a sufficiently high degree of sharpness, which is similar to the degree of sharpness of the required ghost projection.
    \item Sub-speckle size resolution features can be created using ghost projection, at the expense of an increased pedestal, provided the mask has a non-negligible degree of power-spectrum signal at the associated length scales.
    \item The larger the number of available random masks, the better the results of the ghost projection, with respect to quality metrics such as signal-to-noise ratio. However, the larger the set of available masks, the greater the computational expense of selecting which masks to employ.
    \item The ghost-projection signal, as described here, is equal to the image norm $\sqrt{\text{E}[I^2]}$. For self-similar images, enlarging the object approximately linearly increases the final SNR. For general objects, there appears to exist a looser, positive correlation between increasing $\sqrt{\text{E}[I^2]}$ and final SNR. This implies that zooming in as much as is reasonably possible on the features that one wishes to ghost project should improve final SNR.
    \item A trade-off needs to be struck, balancing (i) an accurate spatial resolution of the desired image and random speckle masks, against (ii) the benefits of a coarse spatial resolution such as reduced computational expense and fewer filtered masks $N'$. If such a compromise is difficult, one may be able to resort to combined raster scanning and ghost projection, where we isolate a small window in which to ghost project the desired features (i.e.~projecting the desired image as a collection of tiled projections). 
    \item Using two or more independently-translatable consecutive masks exponentially increases the number of random-mask configurations available, and can improve their characteristics in terms of making them higher in contrast and increasing the average overall absorbency.
\end{enumerate}

Taking on board these recommendations, we can perform a simulation for the more optimal conditions of (i) employing consecutive masks, with (ii) $50nm$ randomly sampled FOVs, for (iii) the slightly Gaussian smoothed binary resolution chart ($\sigma = 0.5$ pixels). This yielded: a NNLS SNR of $3.92 \times 10^7$ with a pedestal of 33.5, $N' = 1\,514$. Adding in Poisson, exposure and translational noise independently, this dropped to an SNR of $7.36 \pm 0.14$ for $\lambda = 10\,000$, $9.50 \pm 0.72$ for $\sigma_w = 1/100$ and $11.36 \pm 0.49$ for perturbations with $\sigma_{ij} = 1/10$. Finally, once all types of noise were included, the NNLS SNR reduced to $5.11 \pm 0.19$. 

Next, we can attempt to trade some of the original NNLS SNR for a reduced pedestal which will hopefully improve the noise robustness of the ghost projection and improve the final ghost projection SNR. Increasing the number of sampled random masks to $100nm$ and enforcing a pedestal of 10, we obtain a NNLS SNR of 15.60 with $N' = 1\,093$. Including Poisson, exposure and translational noise independently produced the SNR results of $10.21 \pm 0.17$, $9.22 \pm 0.44$ and $12.77 \pm 0.12$, respectively. Including all forms of noise produced a final SNR of $7.17 \pm 0.26$.  This final SNR is a 2.7 times increase on our baseline SNR value of $2.67 \pm 0.11$ achieved with $5nm$ Ni foam masks of the binary resolution chart, Fig.~\ref{fig:six graphs}.

\section{Discussion}

This discussion is broken into two parts.  Section \ref{sec:DiscussionA} explores a number of practical questions and potential avenues for future work.  Section \ref{sec:DiscussionB} discusses several theoretical and conceptual questions. 

\subsection{Practical considerations}\label{sec:DiscussionA}

We have mentioned, on several occasions, the tradeoff between (i) the experimental benefits of having a larger set of candidate masks from which to select, for the purposes of ghost projection, using e.g.~the idea of two or more consecutive independently-translatable masks, and (ii) the increased computational effort needed to choose which particular mask subset to employ.  Having more masks to choose from is advantageous since it may, and in general will, enable the ghost projection to be more efficient, provided the appropriate filtration scheme is found. Two practical issues can be mentioned, in this context.  Firstly, the previously-mentioned tradeoff is essentially one between experimental cost and computational cost.  As computing power increases, and optimization algorithms improve, the additional computational cost mentioned above will become progressively less of a constraint, and progressively better at finding weights $w_k$ that lead to a ghost projection of improved quality at fixed or reduced experimental cost.  Secondly, the increased computational effort (as mentioned above) may in principle be decoupled entirely from associated experimental costs that relate e.g.~to experiment duration and experiment complexity.  For example, whether it takes a microsecond or a month of computing time to find an efficient set of mask-illumination weights $w_k$, the resulting experimental ghost-projection exposure time will be the same.  Moreover, if one wishes to make multiple ghost projections of the same pattern, for example in a manufacturing context whereby the ghost projection illuminates a lithographic substrate, then the mask-illumination weights need only be calculated once.

Another interesting practical question concerns the use of a shutter in experimental ghost-projection protocols.  In principle, the need for a shutter can be eliminated by transversely scanning the mask or masks in a continuous manner, while continuously exposing them to the illuminating matter or radiation wave field.  While the resulting ghost-projection quality metrics are likely to be inferior to those which can be obtained by additionally incorporating a shutter (as described in preceding sections), shutter-free schemes might be more realistically achievable in experiment, compared to ghost-projection schemes that incorporate a shutter.  Shutter-free schemes might be of particular utility when using matter or radiation wavefields for which the shuttering process may be impracticably or unacceptably slow for the purposes of ghost projection.  When a single mask is employed, the required shutter-free ghost projection will be encoded in a continuous transverse mask-displacement trajectory
\begin{equation}\label{eq:ParamterisedPath}
    \mathscr{P}(t)=(\delta x(t),\delta y(t)), \quad t_i \le t \le t_f,
\end{equation}
in the $(x,y)$ plane perpendicular to the illumination direction.  Here, the time $t$ parameterizes the mask-displacement path, which is traversed between the initial time $t=t_i$ and the final time $t=t_f$. This mask-displacement trajectory forms a continuous generalization of the ``traveling salesperson path'' illustrated in Fig.~\ref{fig:TSP}a.  The  ghost-projection encoding in Eq.~(\ref{eq:ParamterisedPath}) includes both the mask-displacement path itself, and the speed 
\begin{equation}
    s(t)=\sqrt{\left[\frac{\mathrm{d}}{\mathrm{d}t}\delta x(t)\right]^2+\left[\frac{\mathrm{d}}{\mathrm{d}t}\delta y(t)\right]^2}, \quad t_i \le t \le t_f
\end{equation}
at which each point on the path is traversed.  Additional simplicity of implementation can arise from traversing this mask-displacement path at constant speed.  These and similar shutter-free ghost-projection scenarios could be incorporated into suitably modified optimization schemes, compared to those given earlier in the present paper. Shutter-free schemes could also be readily adapted to the case of two or more consecutive masks.   

Next, we further consider the computational optimization inherent in all ghost-projection schemes considered here.  We have used one of the simplest approaches to computational optimization, namely NNLS, but more sophisticated optimization schemes could also be employed to choose the mask weights $w_k$.  This choice of optimization method is likely to be of particular practical significance when using two or more consecutive masks, on account of the increased computational cost associated with an extremely large number of possible mask configurations.   For example, if one employs two consecutive masks that can each be translated by $2 \, 000$ steps in two transverse directions, there will be $(2 \, 000)^4\approx 2\times 10^{13}$ candidate masks, from which only a very small fraction need be selected for the purposes of efficient ghost projection.  Three such masks would give $(2 \, 000)^6\approx 6\times 10^{19}$ candidate configurations. Computational optimization is such high-dimensional parameter spaces would likely benefit from more sophisticated  approaches than we have considered.  In this case the NNLS-derived weights in our paper could be viewed as providing an already-feasible baseline that has significant room for improvement.  Returning to the above two-mask example, one could specify that $10 \, 000$ masks be selected, which amounts to using only $1$ out of every $2\times 10^9$ available two-mask configurations.  Evidently, the vector of weights $w_k$ will be sparse \cite{Donoho2006,Rani2017} in the sense that only $1$ out of every $2\times 10^9$ coefficients is non-zero.  The high-dimensional nature, of the parameter space associated with the available masks, is likely to give considerable scope for improving the efficacy and practicality of ghost projection, beyond what has been achieved in the present paper.

We close this subsection with some miscellaneous remarks:

\begin{enumerate}

\item The transverse positioning reproducibility, of the ghost-projection mask or masks employed in our method, needs to be sufficiently high.  Roughly speaking, this positioning reproducibility should be the smaller of the following two quantities: (i) the desired resolution of the ghost projection, and (ii) the characteristic transverse length scale of the speckles that are produced by the random mask. 

\item Optimization may be prone to over-fitting noise or discretization artifacts present in the measured masks that may not be present in a more accurate and precise measurement of the mask. That is, it may be hard to distinguish fluctuating noise occurring in one measurement from the permanent random ``noise'' of the mask and one should be careful to remove the former contribution of noise from the latter when forming the random masks. 

\item  Ghost projection formed via NNLS may be the optimal random mask representation, but is not the representation that is most robust to noise inclusions. An avenue for future work could be to write an efficient optimization algorithm that has input parameters for noise inclusions and finds the random mask representation that achieves the best final SNR, as opposed to best initial SNR. 

\item  The exposure noise and translational noise can be reduced from the contributions considered here, whereas Poisson noise and, to some degree, the random mask reconstruction noise, are inherent to optical ghost projection. 

\item A simple and practical way to achieve region of interest (ROI) ghost projection might be to have an additional mask or masks which serve the purpose of a transversely-movable aperture or diaphragm.  For example, having one square aperture of spatial dimensions $A \times A$ (in addition to the speckle-generating mask or masks) could allow a field of view with this spatial extent to be written anywhere within a significantly larger field of view.  Alternatively, having two such independently-translatable square apertures could allow rectangular fields of view with tunable spatial extent $B \times C$ to be written, with $0 < B \le A$ and $0 < C \le A$.     

\item Section~\ref{Old AppB} points out a connection between the traveling salesperson optimization problem, and the choice of a scanning protocol for single-mask ghost projection.  If we instead employ two consecutive masks for ghost projection, we have a two-salesperson analogue of this classic optimization problem, whose solution addresses the practical question of minimizing the experimental exposure time in two-mask ghost projection.

\item We implicitly assumed the total illumination time of the ghost-projection mask to be  sufficiently short that the mask is not damaged to any appreciable degree.  If the damage threshold for the mask is exceeded, the mask will need to be re-characterized or replaced with another previously-characterized mask, for the ghost projection to proceed.

\item Care should be taken to ensure that the ghost-projection illumination plane is only exposed when the mask is an any one specified position, with no exposure being registered when the mask is being transversely moved between respective positions.  A simple way to achieve this is by using a suitable shutter, via the following protocol: 
\begin{enumerate}
\item with the shutter closed, the mask is transversely displaced to a specified position; 
\item the shutter is then opened, and the specified mask is illuminated for a specified time; 
\item the shutter is then closed; 
\item if more exposures are required, return to the first step in the protocol, otherwise the ghost projection is complete.  
\end{enumerate}
Note, also, that allowing for exposure while moving the mask would be practical and is a subject for future research.

\item At the end of Sec.~III, we briefly mentioned the concept of propagation-based phase contrast \cite{Bremmer1952,Snigirev1995,Cloetens1996,Wilkins1996,Paganin2006,KleinOpat1976}, associated with free-space diffraction from the exit surface of the random mask to the ghost-projection plane $Q$. Let us briefly expand on this point.  For small propagation distances where the Fresnel number \cite{SalehTeichBook} is appreciably greater than unity, and for fields with  sufficiently high spatial coherence (e.g.~via a sufficiently small source), propagation-based phase contrast enhances fine spatial detail in a registered intensity distribution \cite{Wilkins2014}.  This boosting of fine spatial detail is related to the fact that the measured contrast is proportional to the transverse Laplacian of the pre-propagation phase map (wave front) of the propagating field \cite{Bremmer1952,CowleyBook,Wilkins1996}. The fact that propagation-based phase contrast preferentially amplifies high-spatial-frequency content, in the transverse intensity distribution created by a ghost-projection mask, may be compared to the complementary case of an absorption-contrast mask (which preferentially creates low-spatial-frequency contrast). This suggests a second point of utility for a two-mask ghost-projection system, beyond that which has already been explored in the main text. Let $R_1$ be one random mask in a two-mask ghost-projection system (Fig.~\ref{Fig:GenericGhostProjectionSetup}(b)); $R_1$ is assumed to be an absorption-only mask, e.g.~by having a sufficiently small distance between the exit surface of $R_1$ and $Q$. Low-spatial-frequency content in the corresponding ghost projection $P$ can be provided by $R_1$.  Conversely,  high-spatial-frequency content in $P$ can be provided by a refractive phase-contrast mask $R_2$ placed at some distance upstream of $Q$. This second mask will generate high-spatial-frequency speckles due to propagation-based phase contrast, if the degree of spatial coherence is high enough. Taken together, $R_1$ and $R_2$ may enable a fuller coverage of the required spatial-frequency range to be accessed, for the purposes of ghost projection.  Moreover, the spatial-frequency content associated with $R_2$ may be tuned by altering the distance between the exit surface of $R_2$ and $Q$.  This may lead to some degree of tuning that can be achieved, for the intensity power spectrum of $P$ that is produced by a two-mask ghost projection system.  Interestingly, in employing propagation-based phase contrast to sharpen the speckle associated with a spatially-random refractive mask, the proximity-correction effect alters from being a difficulty \cite{Bourdillon2000,Bourdillon2001} to an enabler, i.e.~the proximity effect improves rather than degrades the quality of a ghost-projection system, at least in principle. 

\end{enumerate}

\subsection{Theoretical and conceptual considerations}\label{sec:DiscussionB}

Orthogonality of a mask basis does not imply the optimality of that basis for the purposes of ghost projection.  Of course, if one has access to a complete set $\mathcal{B}_1$ of $N$ orthogonal masks,\footnote{Here, ``complete'' means ``complete up to a specified spatial resolution''.  While an infinite number of masks would be needed for the set to be truly complete, when working to a specified spatial resolution, we speak of finite-member mask set as complete if any image can be projected using that basis, to the specified resolution.} any desired distribution of radiant exposure may be projected as a linear combination of members from $\mathcal{B}_1$.  Turning to an overcomplete set $\mathcal{B}_2$ of non-orthogonal masks, such as the ensemble of ``speckle masks'' that has featured prominently in the present paper, it would be incorrect to consider $\mathcal{B}_2$ to be sub-optimal for the purposes of ghost projection, relative to $\mathcal{B}_1$, for the putative reason that $\mathcal{B}_2$ has more members than $\mathcal{B}_1$.  This is related to the possibility of sparse ghost-projection representations, in the sense of the term ``sparse'' employed in compressive sensing \cite{Donoho2006,Rani2017}.  An extreme limit-case example illustrates our main point.  Basis $\mathcal{B}_1$ might consist e.g.~of a set of $N$ identical-shape pinholes of diameter $D$, with each pinhole center being located at one distinct point on the lattice
\begin{equation}
(x,y)=(mD,nD)\in\Omega,
\end{equation}
where $(m,n)$ are integers and $\Omega$ is a finite rectangular region $|x|\le a,|y|\le b$ of width $2a$ and height $2b$.  Alternatively, $\mathcal{B}_1$ might consist of the set of $N$ two-dimensional Fourier harmonics 
\begin{equation}
W_{m,n}(x,y)=\exp(2\pi i m x/a)\exp(2\pi i n y / b) 
\end{equation}
over $\Omega$, each of which have radial spatial frequencies
\begin{equation}
k_r=\sqrt{(m/a)^2+(n/b)^2}
\end{equation}
that do not exceed the Nyquist limit \cite{Press2007} \begin{equation}
\kappa=\tfrac{1}{2}D^{-1}  
\end{equation}
corresponding to a specified spatial resolution $D$. Conversely, let basis $\mathcal{B}_2$ consist of an arbitrarily large number $\tilde{N}\gg N$ of realizations of a certain stochastic process, that yields spatially statistically stationary speckle fields over $\Omega$, whose Fourier power spectra have radial spatial frequencies $k_r$ that are only non-zero over the disc $k_r \le \kappa$.  In principle, every possible ghost projection at resolution $D=\tfrac{1}{2}\kappa^{-1}$ is a member of $\mathcal{B}_2$, therefore the ghost projection can be performed using a single mask drawn from the countably-infinite set $\mathcal{B}_2$. Interpolating away from this unrealistic extreme-case example, towards a practical ghost-projection strategy, we are led to ``compressive ghost projection'': rather than exposing just one mask from $\mathcal{B}_2$, one can instead consider a small subset of masks drawn from $\mathcal{B}_2$, with this subset being a function of the desired ghost projection.  Here, ``small'' means ``having less than $N$ members''. This subset is chosen to enable a sparse representation of the desired ghost projection, which corresponds to being able to perform the ghost projection with fewer than the $N$ masks that would be required of a complete orthogonal basis. The larger the cardinality $\tilde{N}$ of the basis set $\mathcal{B}_2$ from which we can select, the sparser the representation can likely be.  It is for this reason that we have emphasized, earlier in the paper, the use of two of more consecutive masks to significantly increase the number of candidate masks for the purposes of ghost projection.

The preceding paragraph leads us to compare (i) the question of illuminated-sample dose reduction in ghost imaging, to (ii) the related question of minimizing the number of exposed masks for the purposes of ghost projection.  Recall that, in the absence of {\em a priori} knowledge regarding the sample, a classical computational ghost-imaging (CCGI) strategy employing non-orthogonal masks will in general give a dose to the illuminated sample that is no lower, and is often typically higher, than CCGI employing orthogonal masks \cite{Gureyev2018,ceddia2018random,LaneRatner2020,Kingston2021}.  All of the just-cited studies conclude that a necessary condition, for CCGI using non-orthogonal masks to reduce dose in comparison to CCGI using orthogonal masks, is for one to have suitable {\em a priori} knowledge regarding the sample.  This condition is necessary but certainly not sufficient, with work currently ongoing to seek regimes in which {\em a priori} knowledge may be meaningfully leveraged to achieve dose reduction in CCGI.   Turning to ghost projection, there is the crucial point of difference that we have total prior knowledge of the image that we wish to ghost project.  It is this knowledge that enables us to progressively reduce the number of masks that need to be employed for the ghost projection of a specified distribution of radiant exposure, as the number of masks is progressively increased, in the overcomplete set of masks from which we can select a suitable subset.

Another possible extension of the method is that we can replace the projection plane in Fig.~\ref{Fig:GenericGhostProjectionSetup}, with a curved surface $\mathcal{Q}$.  This curved surface should be such that every straight line, radiating from the center of the illuminating source, intersects $\mathcal{Q}$ no more than once.  The ensemble of illuminated masks generates an ensemble of intensity distributions over $\mathcal{Q}$, which can still be used as a basis from which to create a desired ghost projection over $\mathcal{Q}$.  Both the spatial resolution and SNR, of the ghost projection, will be a function of position over the surface $\mathcal{Q}$.

It is also worth discussing the idea that the setup conditions for speckle generation can significantly alter the statistical properties of the speckle fields that are generated for the purposes of ghost projection.  For example, if an optically-thin spatially random mask is coherently illuminated, the nature of the speckle measured at the exit surface of the mask (contact plane) will be significantly different from the nature of the speckle measured in the contact plane for an optically-thick mask.  In the former case, the projection approximation may be employed, whereas in the latter case a fully dynamical scattering theory such as dynamical x-ray diffraction or x-ray multi-slice would be required \cite{Paganin2006}.  More importantly, in the former case the speckle would not be `fully developed', in the sense of the term that is commonly employed in coherent optics, whereas in the latter case the speckle would be fully developed if the screen were to be sufficiently thick.  Even though, as shown by this example, the setup conditions would significantly change the properties of the speckle field generated by a particular spatially-random mask in the context of ghost projection, our method is largely independent of the specific statistical properties of the detected speckle. Broadly speaking, (i) the higher the contrast of the speckle, the higher the contrast of the ghost projection will be, and (ii) the higher the spatial resolution of the speckles, the higher will be the maximum achievable spatial resolution of the ghost projection.

Finally, we discuss conceptual connections of the ghost-projection method with two other approaches in physical optics.  (i) As mentioned in the introduction, ghost projection may be viewed as a reversed form of classical computational ghost imaging \cite{paganin2019writing,ceddia2022ghost}.  (ii) There are also some evident connections with the intensity interferometer of Hanbury Brown and Twiss (HBT) \cite{HBT1,HBT2,HanburyBrownBook}, which is widely used in astronomical imaging.  In the intensity interferometer, one measures intensity correlations in order to image an unknown object, often in the presence of a model for that object, as is the case in using intensity interferometry to measure the angular diameter of stars \cite{HBT1,HBT2,mandel1995optical}.  Under this view, classical computational ghost imaging \cite{shapiro2008computational} may be viewed as a parallelized form of HBT-type intensity interferometer \cite{HBT3,Kingston2020}.  The analogy with ghost projection then becomes clear: ghost imaging measures HBT-type intensity correlations in order to image an unknown transmission function, whereas ghost projection establishes such intensity correlations to create a known transmission function upon temporal integration.

\section{Conclusion}

Ghost projection is the process of shaping a pattern of radiant exposure via a series of random masks. Using experimentally imaged random masks in the x-ray regime, namely a metallic slab of Ni foam and a sheet of 120 grit sandpaper, we explored realistic ghost projection by means of simulation. The realistic noise inclusions considered here were: (i) Poisson noise applied to the physical process of photon counting; (ii) exposure noise, in the form of a Gaussian perturbation, applied to the desired exposure time; (iii) translation noise, also in the form of a Gaussian perturbation, applied to the transverse positioning of the random masks; and (iv) non-uniform illumination in the form of a Gaussian transmission profile. In a noise-free environment, we found that the experimentally acquired random masks were capable of shaping any desired distribution of radiant exposure to a near perfect degree. With noise inclusions, and using experimental parameters taken from commercially available equipment, we found that it is reasonable to ghost project a desired image in the x-ray regime with a resolution on the order of micrometers, to a SNR on the order of 10 with a photon contrast of $10\,000$ and a pedestal of uniform exposure measuring $100\,000$ photons. Depending on the source flux and shutter speed tolerance, we estimated that performing such a procedure may take on the order of minutes to an hour. 

The values obtained here regarding the performance of ghost projection bode well for practical implementation but by no means represent the best that ghost projection may produce. We have outlined how ghost projection can be improved by the tuning of mask parameters, such as the average transmission value, the variance of transmission values and the mask Fourier power spectrum. Furthermore, we also suggested that the numerical optimization algorithm employed could be a source of significant improvement. That is, for the simulations of ghost projection presented here, we numerically optimized the random mask representation of the desired image without consideration of noise robustness. Ideally, the numerical optimization algorithm would find that representation which is most robust to noise inclusions too. Not only might we want to numerically optimize for noise robustness, but for other constraints such as: (i) minimizing the number of masks employed; (ii) minimizing or placing an upper limit on the experiment duration; or (iii) seeking a continuous transverse scan path of the random mask(s) that removes the need for a shutter and may also be a source of experimental speed up.

The process of ghost projection is universal in its ability to create any desired distribution of radiant exposure and frees us from requiring a precisely configurable optical element. It does come at the cost, however, of depositing a uniform pedestal of radiant exposure. Nonetheless, we still foresee future applications of ghost projection in areas where problems associated with proximity correction or mask manufacturing render existing techniques as infeasible. Ghost projection may find future utility in areas such as: (i) beam shaping hard x rays, gamma rays, or matter wave-fields; (ii) in performing universal-mask-based lithography for electromagnetic waves of energy beyond extreme-ultraviolet radiation; or (iii) in the complementary field of three-dimensional printing via reverse tomography (volumetric additive manufacturing \cite{Beer2019,TomographyInReverse2019}).

\section*{Acknowledgments}

We acknowledge useful discussions with Alaleh Aminzadeh, Wilfred Fullagar, Kaye Morgan, Glenn Myers, Lindon Roberts and Imants Svalbe. We thank the European Synchrotron for granting beamtime on Beamline ID19. DC acknowledges funding via an Australian Postgraduate Award. AMK, DP and DMP acknowledge funding via Australian Research Council Discovery Project ARC DP210101312.

\bigskip

\appendix*

\section{Analytical Calculation of Translational Noise} \label{AppA}

\begin{figure}[ht!]
\centering
\includegraphics[width=0.75\columnwidth]{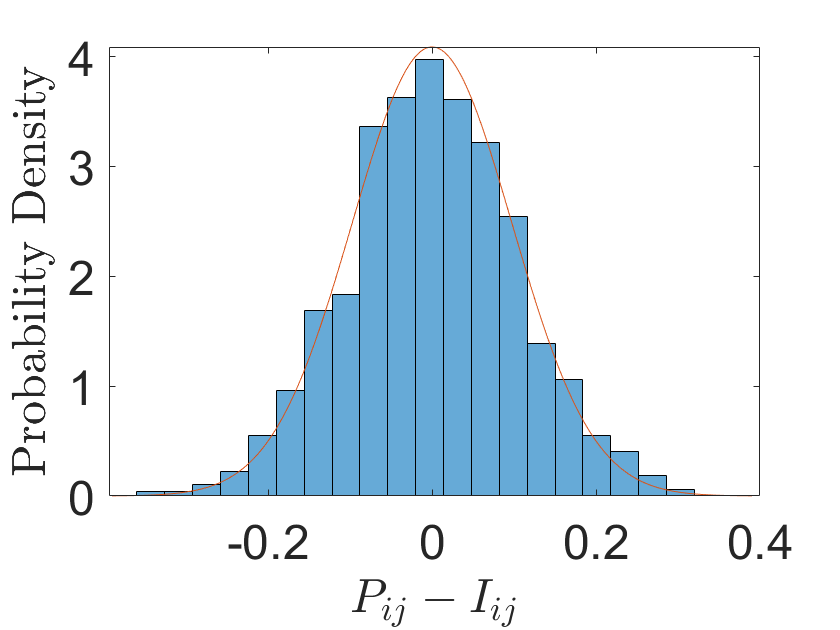}
\caption{Simulated translational noise (see Fig.~\ref{subfig: GP5} for simulated ghost projection) overlaid with the analytically predicted effect of translational noise on a ghost projection with $\sigma_{ij} = 1/10$ and starting NNLS SNR of $3.34 \times 10^8$.}
\label{Fig:Analytical Validation}
\end{figure}

Suppose we have a ghost projection that, in the absence of experimental noise, achieves an effectively perfect reconstruction
\begin{align}\label{Eq:GhostProjectionAppendA}
    P_{ij} = w_k R_{ij}^{ \ \ k} = I_{ij} + \bar{P}.
\end{align}
Adding translational noise to the random masks, we can approximate the perturbed random masks by their first order Taylor series expansion
\begin{align} \label{Eq:PerturbedMask}
    \tilde{R}_{ijk} \approx R_{ijk} + \Delta x_k \partial_i R_{ijk} + \Delta y_k \partial_{j} R_{ijk}, 
\end{align}
where $\Delta x_k$ and $\Delta y_k$ are the $k{\text{th}}$ realization of the $x$ and $y$ translational perturbations and $\partial_i$ and $\partial_j$ are the partial derivatives with respect to the $x$ and $y$ indices (as opposed to the physical partial derivatives). Substituting the expression for the perturbed mask (Eq.~(\ref{Eq:PerturbedMask})) into the expression for the ghost projection (Eq.~(\ref{Eq:GhostProjectionAppendA})), and taking the variance, we obtain
\begin{align}
    \text{Var}[P_{ij}] &= \text{Var} \left[ w_k \tilde{R}_{ij}^{ \ \ k} \right] \nonumber \\
    &\approx \text{Var} \left[  \Delta x_k w_k \partial_i R_{ij}^{ \ \ k} + \Delta y_k w_k \partial_{j} R_{ij}^{ \ \ k} \right] \nonumber \\
    &\approx  \sigma_{ij}^2  w_k^2 \text{Var} \left[  \partial_i R_{ij}^{ \ \ k} + \partial_{j} R_{ij}^{ \ \ k} \right].
\end{align}
Above, $\sigma_{ij}$ is the standard deviation in the translational perturbations. Substituting this into the expression for SNR (Eq.~(\ref{eq:SNRdefinition})), we find the expected SNR of a ghost projection under translational perturbations to be
\begin{align}
    \text{SNR} &\approx \sqrt{ \frac{\text{E}[I^2]}{\sigma_{ij}^2  w_k^2 \text{Var} \left[  \partial_i R_{ij}^{ \ \ k} + \partial_{j} R_{ij}^{ \ \ k} \right]}}  \nonumber \\
    &\approx \frac{\sqrt{\text{E}[I^2]}}{\sigma_{ij}  w_k \mathscr{D}^k},
\end{align}
where $\mathscr{D}^k$ is delroughness (Eq.~(\ref{eq:Delroughness definition})). For the purposes of validation, we can simulate a ghost projection of the binary resolution chart made with $N =5nm$ sampled in strides of 12-by-8 from the Ni foam mask. The comparison of the simulated translational noise and analytically predicted translational noise can be observed in Fig.~\ref{Fig:Analytical Validation}. In terms of SNR, the simulated SNR was $4.52 \pm 0.21$ and the analytically predicted SNR was 4.69.

\vspace{2cm}

\bibliography{GhostProjectionII.bib}

%
%
%

\end{document}